\documentclass[prl,amsmath,amssymb,lengthcheck,showpacs,floatfix,groupedaddress,superscriptaddress,longbibliography]{revtex4-2}
\usepackage{graphicx}
\usepackage[english]{babel}
\usepackage{times}
\usepackage{dcolumn}
\usepackage[usenames,dvipsnames]{color}
\usepackage{units}
\usepackage{bm}
\usepackage{soul}
\usepackage[utf8]{inputenc}

\begin{document}

\title{Subgap states in superconducting islands}

\author{Luka Pave\v{s}i\'{c}}

\affiliation{Jo\v{z}ef Stefan Institute, Jamova 39, SI-1000 Ljubljana, Slovenia}
\affiliation{Faculty  of Mathematics and Physics, University of Ljubljana,
Jadranska 19, SI-1000 Ljubljana, Slovenia}

\author{Daniel Bauernfeind}

\affiliation{Center for Computational Quantum Physics, Simons Foundation Flatiron Institute,
New York, New York 10010, USA}

\author{Rok \v{Z}itko}

\affiliation{Jo\v{z}ef Stefan Institute, Jamova 39, SI-1000 Ljubljana, Slovenia}
\affiliation{Faculty  of Mathematics and Physics, University of Ljubljana,
Jadranska 19, SI-1000 Ljubljana, Slovenia}

\date{\today}

\begin{abstract}
We study an interacting quantum dot in contact with a superconducting
island described by the Richardson model with a Coulomb repulsion term
controlling the number of electrons on the island. This Hamiltonian
admits a compact matrix-product-operator representation and can be
efficiently and accurately solved using the density-matrix
renormalization group. We systematically explore the effects of the
charging energy $E_c$. For $E_c$ comparable to the superconducting
gap $\Delta$, the subgap states are stabilized by the combination of
Kondo exchange coupling and charge redistribution driven by the
Coulomb interaction.  The subgap states exist for both even and odd
superconductor ground-state occupancy, but with very distinctive
excitation spectra in each case. The spectral peaks are not symmetric
with respect to the chemical potential and may undergo discontinuous
changes as a function of gate voltages. 
\end{abstract}

\maketitle

\newcommand{\expv}[1]{\langle #1 \rangle}
\newcommand{\ngs}{{n_{\mathrm{gs}}}}
\newcommand{\nsc}{{n_{\mathrm{sc}}}}
\newcommand{\nimp}{\hat{n}_{\mathrm{imp}}}
\newcommand{\niu}{\hat{n}_{\mathrm{imp},\uparrow}}
\newcommand{\nid}{\hat{n}_{\mathrm{imp},\downarrow}}

The study of long-lived excited states inside the bulk
spectral gap of superconductors (subgap states for short), induced by
impurities and interfaces, drives the development of technologically
important quantum devices. For example, the Yu-Shiba-Rusinov (YSR)
states that result from the exchange interaction which binds a
Bogoliubov quasiparticle at the magnetic impurity site
\cite{yu1965,rusinov1969,shiba1968,sakurai1970} are instrumental in 
realizing topological superconductivity with Majorana edge modes
\cite{NadjPerge2014}.
The excellent understanding of YSR states rests on a theoretical description based
on the Anderson impurity model with a superconducting (SC) bath
described by the Bardeen-Cooper-Schrieffer (BCS) mean-field Hamiltonian
\cite{zmha,satori1992,sakai1993,yoshioka2000}, which can be
tackled using modern impurity solvers
\cite{wilson1975,pillet2013,lee2017prb,Luitz:2010bn,Luitz:2012jz,Domanski2017,Kadlecova2019}.

A recent development are devices with the SC material epitaxially
evaporated on the nanowire hosting the impurity (quantum dot, QD)
\cite{Albrecht2017,vanVeen2018}. The small SC island in these devices
has a considerable charging energy $E_c$
and strong even-odd occupancy effects
\cite{Averin1992,Jank1994,Golubev1994,vonDelft1996} that require an
appropriate description
\cite{Mastellone1998,Braun1998,Dukelsky1999,Dukelsky2000,Braun1999,VanHoucke2006,Rombouts2006}.
This is similar to SC metallic grains
\cite{Tuominen1992,Lafarge1993b,Lafarge1993,Eiles1993,vonDelft2001}
described by the Richardson model, a charge-conserving
Hamiltonian with pairing between the time-reversal-invariant
pairs of states in the orbital basis, which is the appropriate
generalization of the BCS pairing Hamiltonian to a situation with no
translation invariance \cite{anderson1959}. For weak to moderate
pairing and a dense set of levels, the Richardson model is
fully equivalent to a BCS superconductor and has the same low-energy
excitation spectrum, but it is more general: it also describes the
transition to a Bose-Einstein condensate for strong pairing and it
remains applicable for a very small number of levels. 
Without the impurity, the Richardson model can be expressed in
terms of hard-core bosons (paired electrons) and exactly solved via
Bethe ansatz (Richardson-Gaudin equations) \cite{Richardson1963,
Richardson1964, Richardson1966, vonDelft1999, Sierra2000}. The
impurity breaks integrability by splitting the electron pairs through
exchange scattering, thereby precluding this approach. The problem
also cannot be solved using conventional impurity solvers because the
bath is interacting, while the mean-field decoupling of the charging
term leads to incorrect results \cite{suppl}. Furthermore, the
charge-counting trick \cite{Lebanon2003,anders2004} is not applicable
to a gapped spectrum \cite{suppl}. A theoretical tool for this family
of problems has therefore been sorely lacking and some key questions
remained unanswered, in particular whether any states remain present
in the gap for odd occupancy of the SC and, if so, what is their
nature.

Here we show that Richardson-type Hamiltonians with long-range
(all-to-all) interactions coupled to an interacting QD admit a compact
representation in terms of matrix product operators (MPO) with small
$9 \times 9$ matrices and can be efficiently solved without any 
approximations in all parameter regimes using the density-matrix
renormalization group (DMRG) \cite{PhysRevLett.69.2863,
SCHOLLWOCK201196,PhysRevB.78.035116}. Possible extensions include the
capacitive coupling between the QD and the island
\cite{coulomb2arxiv}, the spin-orbit coupling in the SC, the case
of a QD in the junction between two islands, as well as various
multiple-QD problems. 

In this work, we systematically investigate the subgap excitations of
the simplest situation: a single QD coupled to a single SC island. 
The qualitative behavior depends on the ratio of $E_c$ over the SC gap
$\Delta$. For $E_c \lesssim \Delta$, the Kondo coupling
drives the YSR singlet-doublet transition. For $E_c \gtrsim \Delta$, even-odd effects arise from
charge quantization: the occupancy of the SC island
varies in steps of one electron similar to a QD in the
Coulomb blockade (CB) regime. Subgap states are present also for odd
occupancy of the SC, but they disperse very differently compared to
even occupancy. The cross-over $E_c \approx \Delta$ regime shows
complex charging patterns and subgap states with unique properties
that strongly depend on the parity of the number of electrons in the
superconductor. For parameters that are typical of actual devices, the
nature of the subgap states is mixed: it is different from the
prototypical YSR states (large-$U$ limit, $E_c=0$) and
Andreev bound states ($U=0$, $E_c=0$), as well as from the subgap
states of QDs coupled to normal-state Coulomb-blockaded reservoirs.

\begin{figure}
\centering
\includegraphics[width=0.5\columnwidth]{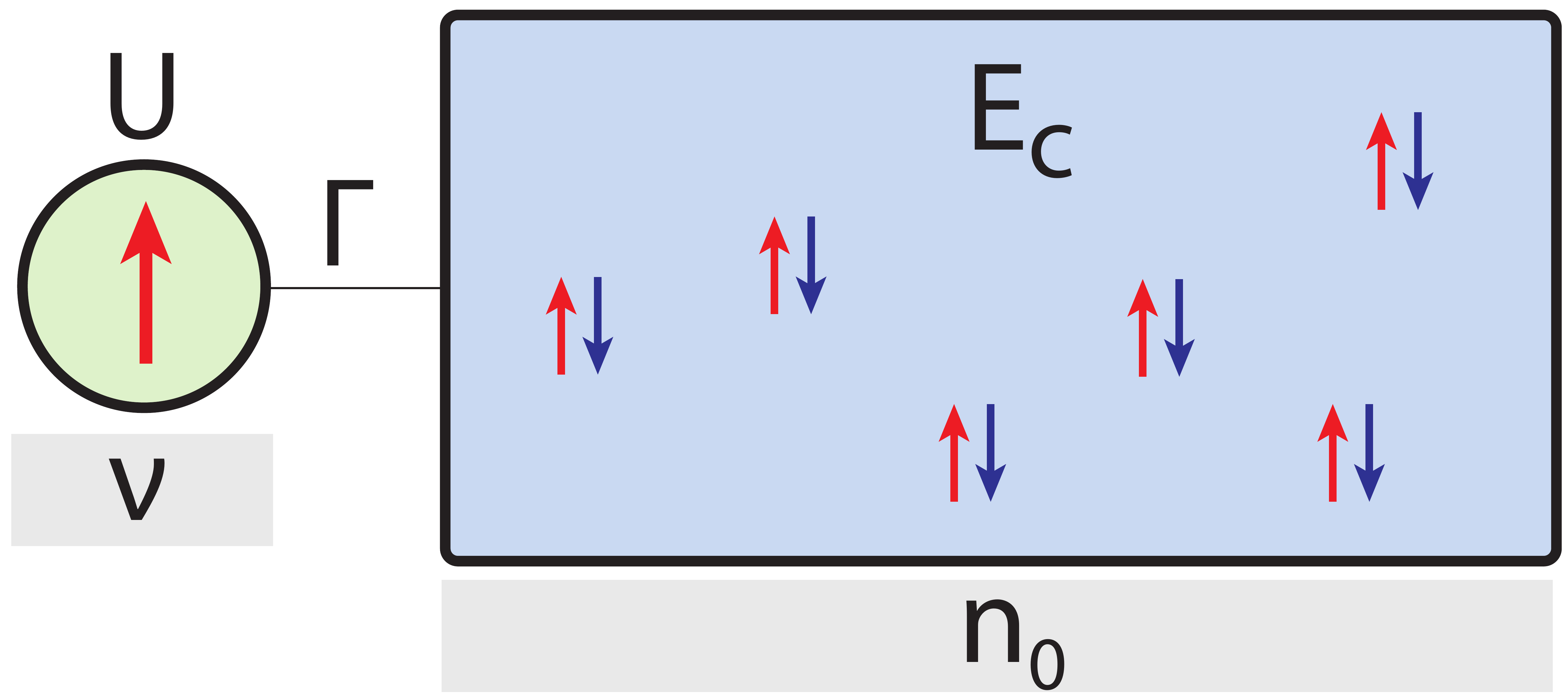}
\includegraphics[width=0.9\columnwidth]{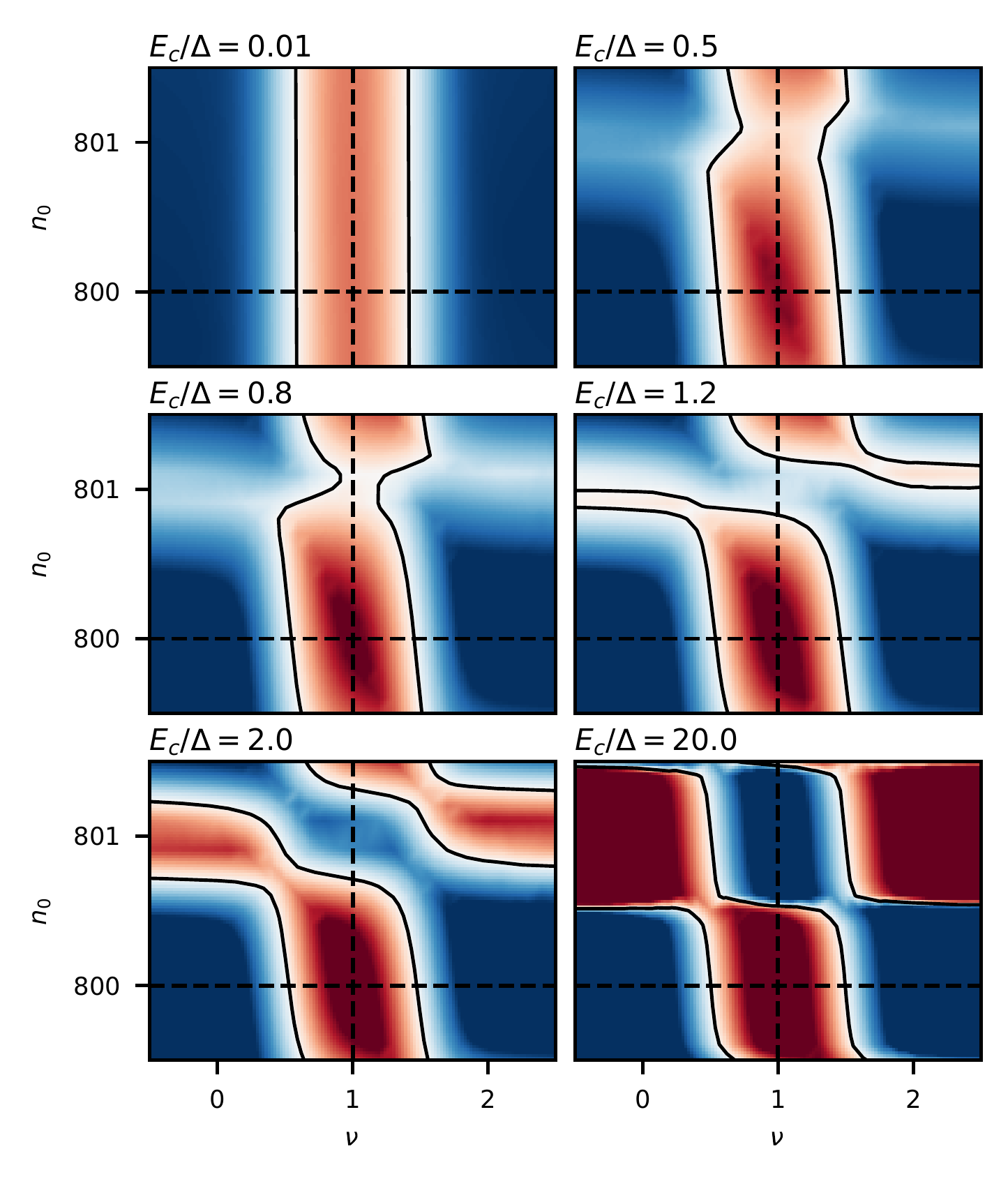}
\caption{Top: schematic representation of the system.
Bottom: phase diagrams as a function of gate voltages applied to the QD ($\nu$)
and to the SC ($n_0$). Dashed lines correspond to half-filling (particle-hole symmetric)
lines of QD and SC at $\nu=1$ and $n_0=N=800$. Red: doublet. Blue:
singlet. The color indicates the energy difference.
}
\label{gate1}
\end{figure}

\emph{Model.} The Hamiltonian we study in the present work is $H = H_\mathrm{imp} + H_\mathrm{SC} +
H_\mathrm{hyb}$ with \cite{anderson1961,Braun1998,Braun1999,vonDelft2001,Dukelsky2004}
\begin{equation*}
\begin{split}
H_\mathrm{imp} &= \epsilon \nimp + U \niu \nid \\ 
&= (U/2) (\nimp - \nu)^2 + \text{const.}, \\
H_\mathrm{SC} &= \sum_{i,\sigma} \epsilon_i c^\dag_{i\sigma} c_{i\sigma}
- \alpha d \sum_{i,j} c^\dag_{i\uparrow} c^\dag_{i\downarrow} c_{j \downarrow} c_{j\uparrow}
+ E_c (\hat{n}_\mathrm{sc}-n_0)^2, \\
H_\mathrm{hyb} &= (v/\sqrt{N}) \sum_{i\sigma} \left( c^\dag_{i\sigma} d_\sigma + \text{H.c.} \right).
\end{split}
\end{equation*}
Here $d_\sigma$ and $c_{i\sigma}$ are the annihilation operators corresponding to impurity and bath, $\sigma=\uparrow,\downarrow$,
$\hat{n}_{\mathrm{imp},\sigma} = d^\dag_\sigma d_\sigma$ and $\nimp=\sum_\sigma
\hat{n}_{\mathrm{imp},\sigma}$.
$\epsilon$ is the impurity level controlled by the gate voltage applied to the QD, $U$ the electron-electron repulsion, and $\nu=1/2-\epsilon/U$ is the
impurity level in units of electron number. The SC has $N$ levels spaced by $d=2D/N$ where $2D$ is the bandwidth,
the orbital indexes $i$ and $j$ range between 1 and $N$, the dimensionless coupling constant for pairing interaction is $\alpha$,
$\hat{n}_\mathrm{sc}=\sum_{i\sigma} c^\dag_{i\sigma} c_{i\sigma}$, and $n_0$ is the gate voltage
applied to the SC expressed in units of electron number.
The hybridisation strength is $\Gamma=\pi \rho v^2$, where $\rho=1/2D$ is the normal-state bath density of states.
A schematic representation of this Hamiltonian is shown in Fig.~\ref{gate1}, top.
Most calculations in this work are performed for $N=800$ and $\alpha=0.23$
(magnitude appropriate for Al grains \cite{Braun1998}), with $D=1$ as the energy unit.
The corresponding gap in the thermodynamic limit is $\Delta \approx \unit[0.026]{D}$ \cite{suppl}.
The interlevel separation is $d=2D/N=\unit[0.0025]{D} \approx \Delta/10$,
thus the finite-size corrections to BCS theory
\cite{vonDelft1996,Mastellone1998,Braun1998,suppl,Roman2003,Yuzbashyan2003}
are relatively small \cite{suppl}. Unless specified
otherwise, the QD interaction
is $U=0.1 \approx 4\Delta$ which is a typical value for nanowire
devices, and $\Gamma=0.1U$ which corresponds to
intermediately strong coupling.
In this work we focus on the situation where the QD-SC system is not
strictly isolated but
in contact with weakly coupled tunneling probes. The ground state (GS) with fixed (integer) total number of
electrons $n=\ngs$ is determined by the gate voltages $\nu$ and $n_0$. We use $(0)$, $(+1)$ and
$(-1)$ as shorthands for the GS and the lowest-energy (subgap) excited states with occupancy
$\ngs\pm1$, respectively.

\begin{figure}
\centering
\includegraphics[width=0.9\columnwidth]{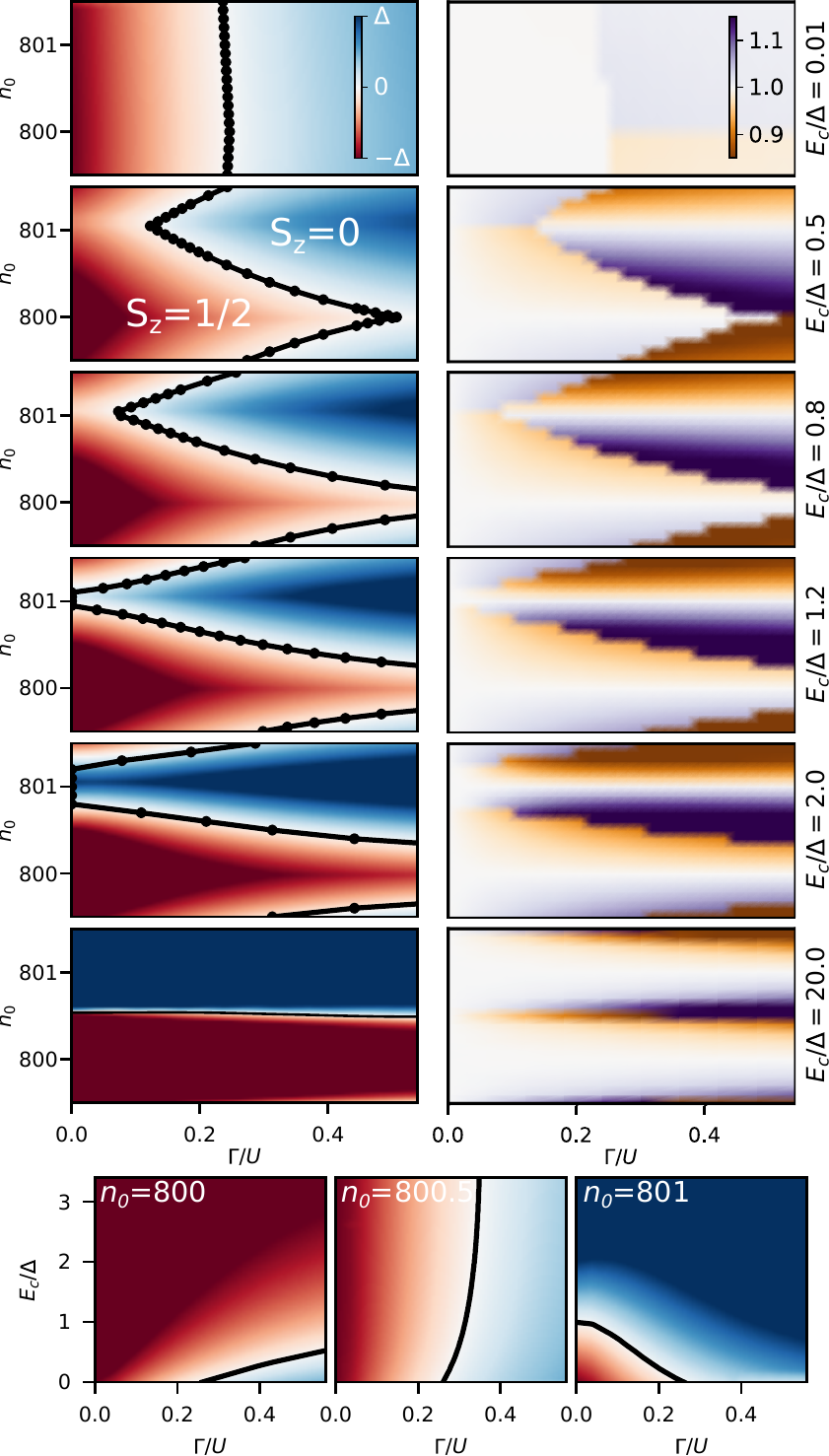}
\caption{Evolution from the YSR regime to the CB regime for $\nu=1$.
Energy difference $E^D-E^S$ between the lowest-lying singlet and
doublet states in the $(\Gamma,n_0)$ plane for a range of $E_c$ (left
panels), and in the $(\Gamma,E_c)$ plane for even, half-integer, and
odd $n_0$ (bottom panels). Red: doublet, blue: singlet, black line:
quantum phase transition. Right panels: QD occupancy variation.}
\label{diag1}
\end{figure}

\begin{figure}
\centering
\includegraphics[width=0.9\columnwidth]{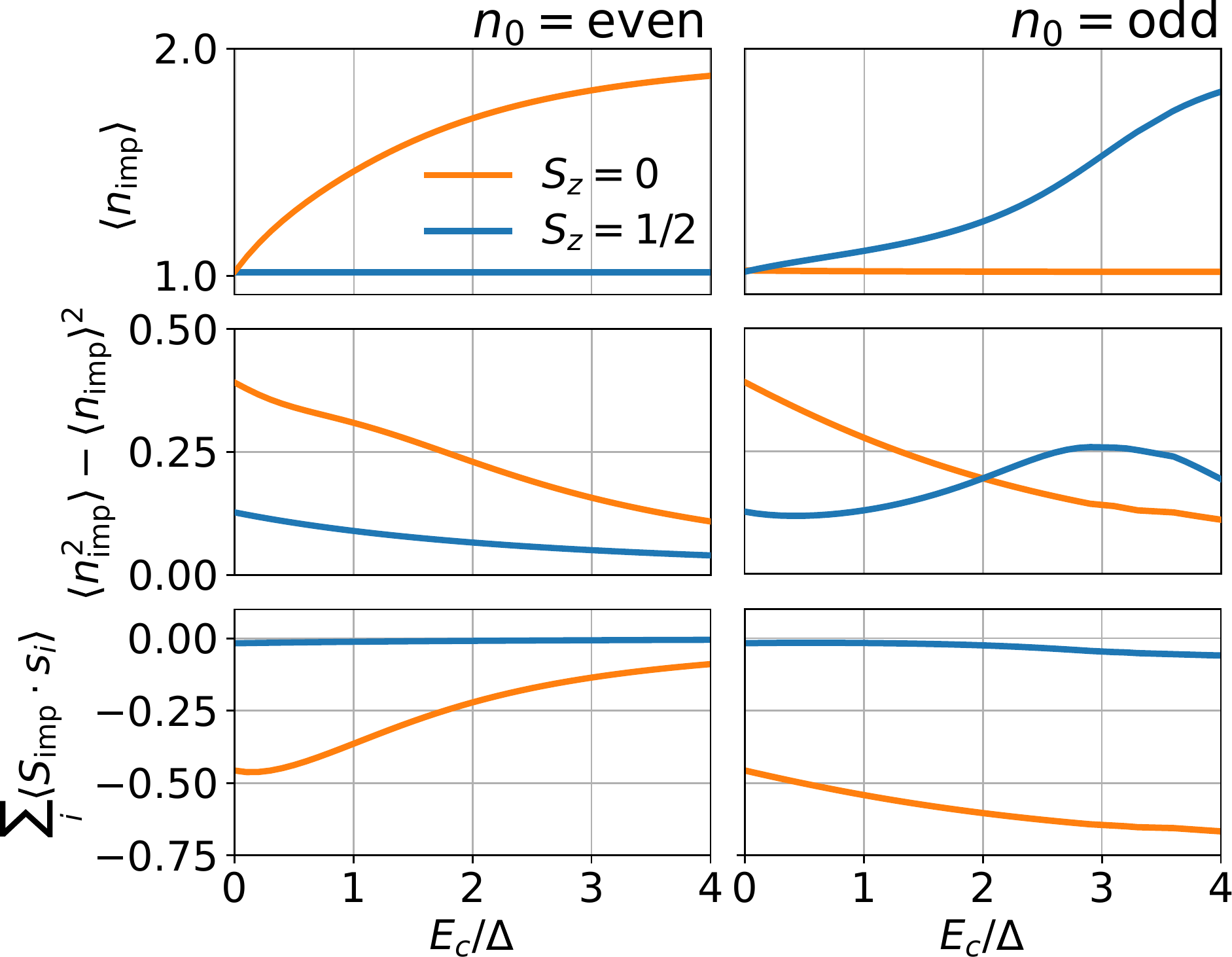}
\caption{$E_c$ dependence of subgap state properties at $\nu=1$.
}
\label{ssnimp}
\end{figure}

\begin{figure*}
\centering
\includegraphics[width=2.0\columnwidth]{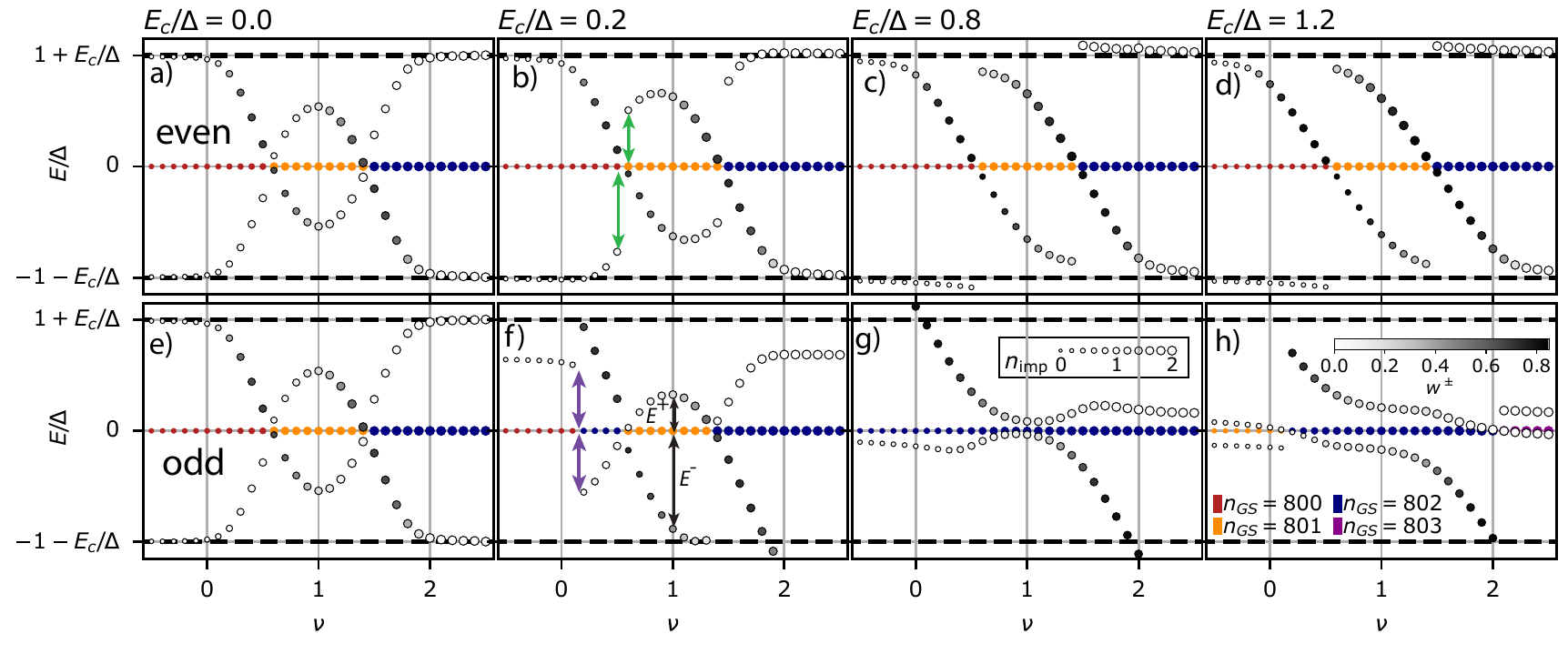}
\caption{Subgap spectral functions for even (upper row, $n_0=800$) and odd integer (bottom row, $n_0=801$) tuning of the superconductor occupancy, as a function of the gate voltage applied on the quantum dot.
The positions $E^+=E^{(1)}-E^{(0)}$ and $E^-=E^{(-1)}-E^{(0)}$ indicate the excitation energies of particle-like (+1) and hole-like
(-1) states, the grayscale shows the corresponding spectral weights $w^\pm$ (see legend in the inset).
The dots at $E=0$ provide information about the ground state: the color encodes the GS charge sector, the size encodes the GS impurity occupancy $n_\mathrm{imp}$ (see legend in the inset).
The size of grey dots denotes the impurity occupancy in the corresponding excited states.
The charge gap for even $n_0$ and $\Gamma\to0$ is $\Delta+E_c$ (dashed lines).
}
\label{fig4}
\end{figure*}

\emph{Results.} The evolution from the YSR to CB regime is clearly
visible in the charging diagrams in the $(\nu,n_0)$ plane, see
Fig.~\ref{gate1}. For small $E_c$ there is a 2$e$ periodicity along
the $n_0$ axis with only a weak even-odd modulation of the subgap
state energies, while for large $E_c$, the system instead shows a
clear 1$e$ periodicity. The transition between the two regimes occurs
gradually for $E_c$ of order $\Delta$, with the charge stability
regions deforming from a pattern of vertical stripes into a
well-defined honeycomb diagram. With increasing $\Gamma$, the singlet
(blue) regions increase in size because the singlet energy decreases
with respect to the doublet energy, while the phase boundaries become
smoother (less rectangular) and develop a diagonal slant because for
large $\Gamma$ each gate voltage influences occupancy in {\sl both}
parts of the system \cite{suppl}.
To better understand this evolution, in Fig.~\ref{diag1} we follow the
dependence on the coupling $\Gamma$ at fixed $\nu=1$, where the QD
hosts a local moment. In the $\Gamma\to0$ limit the impurity is
decoupled and for $E_c < \Delta$ the SC is always in a conventional
BCS state with even $\nsc= \langle \hat{n}_\mathrm{sc} \rangle$. For
$E_c \approx 0$, we uncover the conventional singlet-doublet YSR
transition at $T_K(\Gamma)/\Delta = 0.3$
\cite{satori1992,zitko2015shiba,lee2017prb} for a value of $\Gamma$
that does not depend on $n_0$; here $T_K(\Gamma)$ is the impurity
Kondo temperature at the given value of $\Gamma$.  With increasing
$E_c<\Delta$ the transition point moves to larger values of $\Gamma$
around even $n_0$ where the charging term makes the existence of
Bogoliubov quasiparticles energetically unfavorable. The opposite
holds around odd $n_0$. As $E_c$ grows beyond $\Delta$ we observe a
qualitative change. The SC state in the $\Gamma\to0$ limit now
depends on $n_0$: for $n_0$ close to an even integer value, it is a
BCS state, while for $n_0$ close to an odd integer value, an
additional unpaired electron (Bogoliubov quasiparticle) sits at the
bottom of the quasiparticle band \cite{Averin1992} and for
$\Gamma\neq0$ interacts with the electron at the impurity site via
exchange interaction, forming a singlet GS. The exact location of the
phase boundary depends in a non-trivial way on $\Gamma$, $U$ and $E_c$
due to a three-way competition between Kondo screening, pairing
correlations, and Coulomb interaction.
The latter also leads to a strong charge redistribution between the QD
and SC in the singlet GS, see right panels in Fig.~\ref{diag1}.
Fig.~\ref{ssnimp} shows the $E_c$ dependence of impurity occupancy,
occupancy (charge) fluctuations, and spin correlations at fixed
$\Gamma$, for even and odd SC tuning. As $E_c$ increases, the charge
in the state penalized by the Coulomb term is redistributed, while the
charge fluctuations generally decrease, as expected, except for the
doublet state in the odd-$n_0$ case. The Kondo coupling
decreases/increases with increasing $E_c$ for even/odd parity of
$n_0$, which is reflected in the spin-spin correlations of the
spin-singlet states.

A striking consequence of the Coulomb repulsion is the lack of symmetry in the subgap peak
positions except for special points (e.g. $\nu=1$ and even $n_0$), see Fig.~\ref{fig4}.
This is a significant departure from the conventional case with $E_c = 0$, where the peaks are
{\sl always} located exactly at $\omega = \pm E_\mathrm{YSR}$, so that
the spectra take the form of symmetric eye-shaped loops.
For $E_c\neq 0$, the states $(\pm1)$ have in general different excitation energies $E^\pm$
leading to drastic changes in the spectral shapes even for small $E_c$ (see e.g. black arrows in Fig.~\ref{fig4}f).
In particular, this leads to {\sl discontinuous} changes in the spectrum when the total
occupancy in the GS changes. For instance, as $\nu$ increases, $E^+$ decreases until reaching zero, at
which point the former $(+1)$ state becomes the new GS. At this point the former $(-1)$ is
no longer spectroscopically visible (i.e., it ``disappears''), since it has two electrons less than the new GS. The
same holds for decreasing $\nu$ for $E^-$. An example of such discontinuous changes in the spectrum
is indicated by vertical green arrows in Fig.~\ref{fig4}b.
The spectrum behaves even more remarkably for odd $n_0$ (bottom row of Fig.~\ref{fig4}).
For moderate $E_c/\Delta=0.2$, one observes valence skipping (occupancy jump from 800 to 802,
then back to 801) due to a redistribution of charge between the SC and the QD,
experimentally visible as a two-sided discontinuity (Fig.~\ref{fig4}f, purple arrows),
while for $E_c \lesssim \Delta$ the excitation are pinched at $\nu=1$.
For large $E_c>\Delta$, the spectra eventually transform into straight lines typical of CB systems.

\begin{figure}
\centering
\includegraphics[width=\columnwidth]{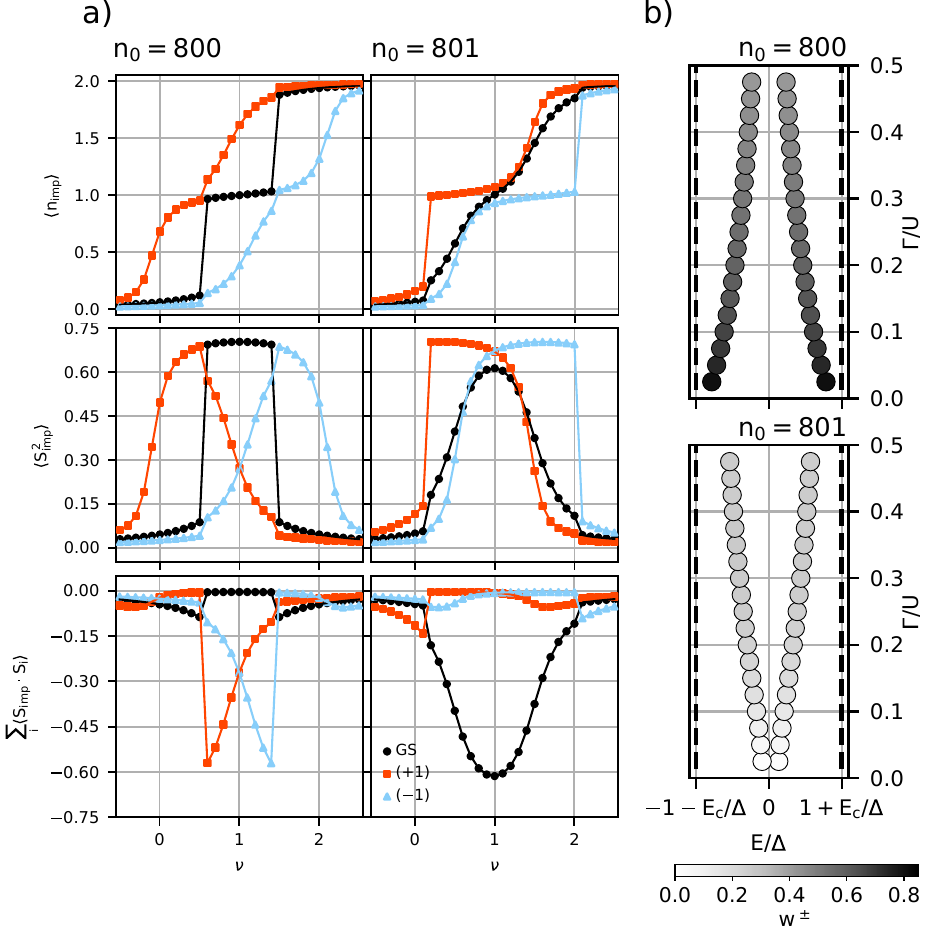}
\caption{Subgap state properties for $E_c/\Delta=1.2$ (right-most panels in Fig.~\ref{fig4}).
a) Expectation values for (0), (+1), (-1) states vs. gate voltage.
b) $\Gamma$ dependence of the subgap spectra at $\nu=1$.
}
\label{figA}
\end{figure}

\emph{Discussion.} 
The nature of the subgap states at $E_c \gtrsim \Delta$ is revealed in Fig.~\ref{figA}, where we show the
properties of (0), (+1) and (-1) as a function of $\nu$ in panel a), and the $\Gamma$ dependence
of the excitations at $\nu=1$ in panel b). We first discuss the case of $\nu=1$. For even $n_0$,
the GS is a decoupled local-moment state, while the states (+1) and (-1) have impurity 
occupancies differing by more than {\it half an electron} compared to the ground state due to the cost $E_c$ of changing the SC
occupancy, but they still carry some local moment that aligns antiferromagnetically with respect
to SC electrons. The excitations detach from the continuum edge at small $\Gamma$ and shift toward
the mid-gap region with increasing $\Gamma$, see Fig.~\ref{figA}b.
The (+1) and (-1) are hence somewhat similar to conventional
YSR singlets, although their impurity local-moment is reduced not only through the Kondo mechanism, but
also by a very large charge transfer to or from the superconductor. 
The excitations may thus be characterized as being YSR-like, sharing some but not all features
of the conventional YSR states at $E_c=0$.
For odd $n_0$, the states (0), (+1), (-1) are all similar to each other and carry a local moment
at the impurity site. They differ mostly in the presence or absence of the lone Bogoliubov
quasiparticle in the SC: (0) has the quasiparticle, while (-1) and (+1) do not; (+1) differs
from (-1) by the presence of one additional Cooper pair. Adding an electron to the GS costs $E_c$ and disrupts the singlet which further costs an energy of order 
$J_K \propto \Gamma$, however at the same time a Cooper pair is formed and the energy $\Delta$ is
gained. Indeed, the results for $n_0=801$ in panel b) show for small $\Gamma$ approximately
linear behavior with a zero intercept at $E_c-\Delta = 0.2 \Delta$. We thus conclude that for odd
$n_0$ the subgap states are doublets which result from the disruption of the strong local QD-SC singlet
formed by the electron in QD and the lone quasiparticle in SC (see also Fig.~S4 for details
\cite{suppl}); these states have no counterparts at all in $E_c=0$ systems.
Fig.~\ref{figA}b shows that for even $n_0$ the excitations have a large weight on the QD, while the opposite is the case for odd $n_0$, where the electron is mostly added to the SC island.

The dispersion ($\nu$-dependence) of excitations for large $E_c$ (e.g.
$E_c \gtrsim 0.8\Delta$) is strongly affected by the charging terms
and it follows the variation of the difference in the impurity occupancy
between the ground and excited states, as revealed by comparing 
Figs.~\ref{figA}a and \ref{fig4}.  For even $n_0$ around half filling,
the GS occupancy is mostly flat as a function of $\nu$, while the (+1)
and (-1) occupancies vary rapidly. This is reflected in an equally
rapid variation of the excitation spectrum at the same parameters.
Along the same line, for odd $n_0$ and away from half filling, the
occupancy of (0) and (+1) are similar for $\nu>1$, while those of (0)
and (-1) are similar for $\nu<1$ and, again, this is reflected in the
spectral shape (in this case as flat sections).

\emph{Conclusion.} Subgap states persist in the presence of large
charging energy but they have properties quite unlike YSR states
\cite{coulomb2arxiv}. The proposed method can also address the
question of quasiparticle poisoning in Majorana islands
\cite{Albrecht2017,Lutchyn2018}, gate sensing of charge-tunneling
processes \cite{vanVeen2019,deJong2019}, superconducting islands on
surfaces \cite{Vlaic2017,Menard2017,Zhang2018,PalacioMorales2019},
topological superconductivity \cite{Lapa2020}, and the existence of
Majorana zero modes beyond mean-field \cite{Ortiz2014}. Generalization
to multiple bands will find application in multi-channel and
topological Kondo effects \cite{zitko2017nfl,Beri2012,Papaj2019}.

\begin{acknowledgments}
We have benefited from suggestions and discussions with Tadeusz
Domansky, Tom\'a\v{s} Novotn\'y, Volker Meden, Marion van Midden,
Andr\'as P\'alyi, Anton Ram\v{s}ak, Roman Rausch, and with the members
of QDev at the University of Copenhagen. R\v{Z} and LP acknowledge the
support of the Slovenian Research Agency (ARRS) under P1-0044.
Calculations were performed with the ITensor library \cite{ITensor}.
\end{acknowledgments}

\bibliography{paper1}

\clearpage

\appendix

\setcounter{figure}{0}
\renewcommand\thefigure{S\arabic{figure}}

\section{SUPPLEMENTAL MATERIAL}

Here we discuss the failure modes of alternative approaches for
solving the impurity problems with superconducting bath and Coulomb
blockade (mean-field decoupling and charge-counting auxiliary
operator approach), show a number of additional results that
supplement those presented in the main text, provide the details on
the method (including a detailed account of the Hamiltonian and its
matrix-product-operator representation), and present some results of
benchmark calculations for method validation.

\subsection{Failure of the mean-field decoupling of the charging term}

The Hamiltonian discussed in this work contains interaction terms of
three types: 1) pairing interaction in the superconductor, 2) on-site
charge repulsion on the impurity site, 3) charge repulsion in the
superconducting island. This raises the question of possible
simplifications of the problem through mean-field decoupling of
certain terms. For problem sizes (number of levels in the SC island)
considered in this work, the mean-field decoupling of the pairing
interaction to the BCS mean-field form $\Delta c^\dag_{i\uparrow}
c^\dag_{i\downarrow}$ is a perfectly valid approximation. In our
scheme, we do not perform this step for purely practical reasons: it
does not simplify the calculations (in fact, the loss of the the total
charge as a conserved quantum number would only make them more
difficult from the numerical perspective). The mean-field decoupling
on the impurity on-site repulsion is well known to lead to
qualitatively incorrect results (e.g., the well-known unphysical
spontaneous spin symmetry breaking in the unrestricted Hartree-Fock
solution of the Anderson impurity model for $U>\pi\Gamma$
\cite{anderson1961}). The remaining question is thus that of the $E_c
(\hat{n}_\mathrm{sc}-n_0)^2$ term in the superconducting bath, whose
mean-field decoupling might at first seem innocuous. If this term
could be safely decoupled without affecting the qualitative nature of
the subgap states and without severe quantitative issues, this would
enable the use of traditional quantum impurity solvers, such as NRG
and CTQMC. 

We thus look into this question more closely and consider two
approximative schemes. One consists in the traditional mean-field
decoupling of this term (the ``mean-field'' scheme), leading to the
following quadratic form:
\begin{equation}
H' =  ( \hat{n}_\mathrm{sc} - n_0 )^2 \approx 2 \hat{n}_\mathrm{sc} (
\langle \hat{n}_\mathrm{sc} \rangle - n_0 ) + n_0^2 - \langle
\hat{n}_\mathrm{sc}
\rangle ^2.
\end{equation}
The problem then needs to be solved self-consistently, for each state
separately (we note that using ground-state expectation values for all
states leads to unphysical discontinuities across the quantum phase
transitions). 

The other scheme (the ``static'' scheme) is a more
severe approximation. It consists in performing the calculations for
$E_c=0$, then simply shifting the resulting energy by $E_c
(\expv{\hat{n}_\mathrm{sc}}-n_0)^2$, again for each state separately.

We note that both the ``mean-field'' and the ``static'' schemes fully
neglect the correlation effects of the $E_c$ term and retain only its
average effect, but they depend in the implementation details: the
first is self-consistent and incorporates the static effect of the
$E_c$ term on the wavefunction, while the second is a one-shot
calculation at $E_c=0$, where the wavefunction does not depend on
$E_c$ and $n_0$ at all, only the energy is shifted.

\begin{figure*}[htbp]
\centering
\includegraphics[width=1.5\columnwidth]{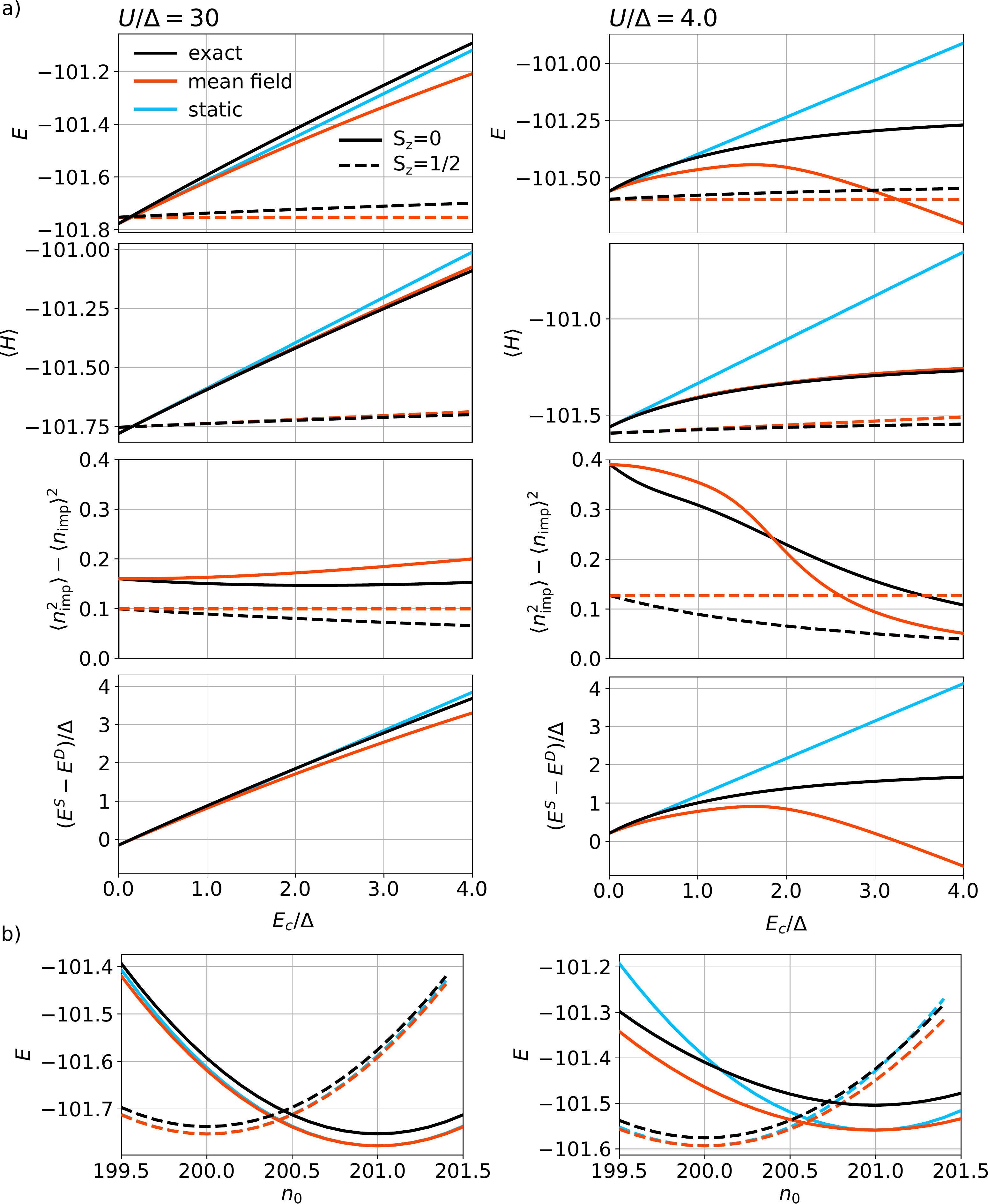}
\caption{Comparison of the mean-field decoupling schemes for large
$U/\Delta$ (Kondo limit) and moderate $U/\Delta$ (realistic value for
actual devices). We compare exact results from the DMRG solution (full
lines) with the results of two approximate schemes (dashes lines). (a)
$E_c$ dependence of energy at fixed $n_0=200$. First row: energies obtained
by the different approaches (see text for details). Second row:
expectation value of the full Hamiltonian evaluated with the
wavefunction obtained in the different approaches. 
Third row: charge fluctuations. Fourth row: excitation energy.
(b) $n_0$
dependence of energy at fixed $E_c=XX$. Parameters are $\Gamma/U=0.2$,
$\alpha=0.4$, $N=200$. } \label{meanfield1}
\end{figure*}

For purposes of comparing these approaches, we performed all
calculations using the same DMRG impurity solver, the only difference
being the different level of the approximation in the charging term.
The $E_c$ dependence at even $n_0$ tuning are shown in
Fig.~\ref{meanfield1}(a). We consider two cases, that of large
$U/\Delta=30$ where the QD behaves as a Kondo impurity to a good
approximation, and that of $U/\Delta=4$ which is appropriate for real
devices. We show two energies. The energy $E$ shown in the first row
is the energy as evaluated within the approximation scheme: this is
the energy that the different methods produce as their final answer.
The expectation value $\expv{H}$ in the second row is computed using
the wavefunctions from the different schemes: this is the actual
energy of the approximate wavefunctions, which can be used as a gauge
for their quality (the exact DMRG solution is the absolute ground
state of the problem with minimal energy).

For $U/\Delta=30$, the differences are only quantitative and
relatively weak. This is expected, since large $U$ enforces small
charge fluctuations on the impurity site, thereby reducing the effects
of the charging term in the range of $E_c$ values of interest (order
$\Delta$). We find that the discrepancy is larger for the singlet
state, where the ``static'' method seemingly performs slightly better
than the self-consistent ``mean-field'' method. For the doublet state,
both approximate schemes produce the same result. This is due to the
symmetry of the problem: the system is particle-hole symmetric in this
case, with $\expv{\hat{n}_\mathrm{sc}}=n_0$. The reason for the energy
difference with respect to the exact solution is the charge
fluctuations, i.e., the difference between
$\expv{(\hat{n}_\mathrm{sc}-n_0)^2}$ and
$(\expv{\hat{n}_\mathrm{sc}}-n_0)^2$, which is non-zero even in the
presence of particle-hole symmetry. In addition, the inexact treatment
of charge fluctuations on the SC island also affects the results for
charge fluctuations on the {\it impurity} site (third row): the
decrease of the impurity charge fluctuations in the doublet state with
increasing $E_c$ is not captured at all at the mean-field level, while
the fluctuations in the singlet state grow rather than slightly
decrease. The plot of $\expv{H}$ (second row) shows that the error
quantified by the difference in energy expectation values is small for
the doublet states, but more significant for the singlet states.
Interestingly, the differences in $E$ are significantly larger than
those in $\expv{H}$ for the ``static'' method, while the
``mean-field'' method shows the opposite. This implies that the
differences in $E$ (i.e., the measurable excitation energies, shown in
the fourth row) are quite small for the ``static'' method, while the
discrepancy is somewhat larger for the ``mean-field'' method. In spite
of all these quantitative differences, we conclude that at this large
value of $U/\Delta=30$, the charging term in the superconductor is for
the most part adequately captured by the mean-field approximations
(which will in fact be used in the following when discussing the
large-$U$ limit, see Eq.~\eqref{Eq1} in the section on phase diagrams
of the Supplemental Material). For such large $U/\Delta$ values, the
states thus still have the nature of the standard YSR states even at
large $E_c/\Delta$, only with shifted energies.

At realistic $U/\Delta=4$, the discrepancies are more severe (see the
right-hand panels). This is especially the case when evaluating the
energies $E$ within the approximate schemes, rather than evaluating
the expectation value of the original Hamiltonian, $\expv{\hat{H}}$.
This holds in particular for the singlet states, which arise from a
strong coupling between the QD and SC electrons. The energies of the
three aproaches start strongly diverging already at low $E_c/\Delta$
(order $0.1$), while at $E_c/\Delta \approx 1$ the error is
sufficiently large that the results are highly questionable. The
mean-field approach fails even qualitatively, predicting a quantum
phase transition which does not really exist. At the same time, the
mean-field wavefunction appears to be reasonably good as far as its
energy expectation value is concerned. The plots of charge
fluctuations reveal that the mean-field methods nevertheless poorly
describe the dynamical aspects of the problem. We thus conclude that
at the experimentally relevant values only the full solution leads 
to acceptable results.

The dependence of energies on the gate voltage $n_0$ at fixed $E_c$
are shown in Fig.~\ref{meanfield1}(b). We again find that the
approximations are adequare in the limit of large $U/\Delta$, while
for realistic $U/\Delta=4$ the deviations become sizable. In
particular, we note that the curvatures of the singlet-state parabolas
(centered at $n_0 \approx 201$) are different. The ``static'' method
does not capture the renormalization of $E_c$. The ``mean-field''
method does, but at the price of a sizable systematic error that is
constant in $n_0$.

We conclude that the treatment of the charging term at the mean-field
level is sufficient to obtain some basic understanding of the overall
trends in the limits of low $E_c$ \footnote{A more precise statement
is that the mean-field approximation is appropriate in the limits $E_c
\ll U$ and $E_c \gg U$, where {\it either} of the two charge repulsion
terms fixes the occupancies in {\it both} parts of the system.}, but
it does not capture the changing nature of the subgap states
(evolution from conventional YSR towards subgap states with Coulombic
nature, characterized by local moment reduction via charge
redistribution), produces spurious qualitative features (phase
transitions which do not actually exist), and is insufficiently
accurate at quantitative level to be of real use in the interpretation
of experiments.

\subsection{Failure of the charge-counting trick}

Impurity models with a Coulomb term $E_c \hat{n}^2$ can be mapped on
an effective model using collective charge operators
\cite{matveev1991,Schoeller1994} and solved using the numerical
renormalization group (NRG) techniques \cite{Lebanon2003,anders2004}.
The idea is to replace the hopping term
\begin{equation}
\sum_{i\sigma} c^\dag_{i\sigma} d_\sigma + \text{H.c.},
\end{equation}
by
\begin{equation}
\sum_{i\sigma} \hat{N}^+ c^\dag_{i\sigma} d_\sigma + \text{H.c.},
\end{equation}
after introducing collective charge operators for electrons on the
island, $\hat{N}=\sum_{m=-\infty}^{\infty} m |m\rangle \langle m |$,
$\hat{N}^\pm=\sum_{m=-\infty}^{\infty} |m\pm 1\rangle \langle m |$,
converting $E_c (\hat{n}_\mathrm{sc}-n_0)^2$ to $E_c (\hat{N}-n_0)^2$,
finally relaxing the constraint $\hat{N} = \hat{n}_\mathrm{sc}$
and thus regarding $\hat{N}$ as an independent degree of freedom. 
This approach is applicable when the dynamics of the collective charge
operator is insensitive to the precise number of conduction electrons
in the bands \cite{Lebanon2003}. This is the case for normal-state
bands, but {\it not} for gapped systems such as superconductors, where
this approach leads to incorrect excitation spectra. We find that the
spectra contain not only additional spurious states that would need to
be projected out (replicas of physical states at higher energies which
do not meet the constraint), but even the expected physical
excitations have incorrect energies. This is a fundamental issue that
does not appear to have a practical solution. For this reason, it
appears unlikely that conventional impurity solvers will ever be
adapted to problems with gapped continuum in the presence of a Coulomb
interaction term.

\subsection{Finite-size effects}

\begin{figure}[htbp]
\centering
\includegraphics[width=0.9\columnwidth]{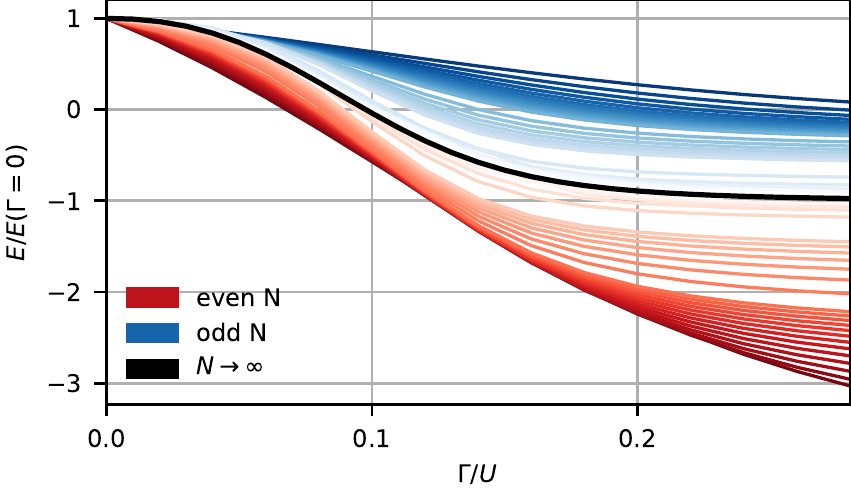}
\includegraphics[width=0.9\columnwidth]{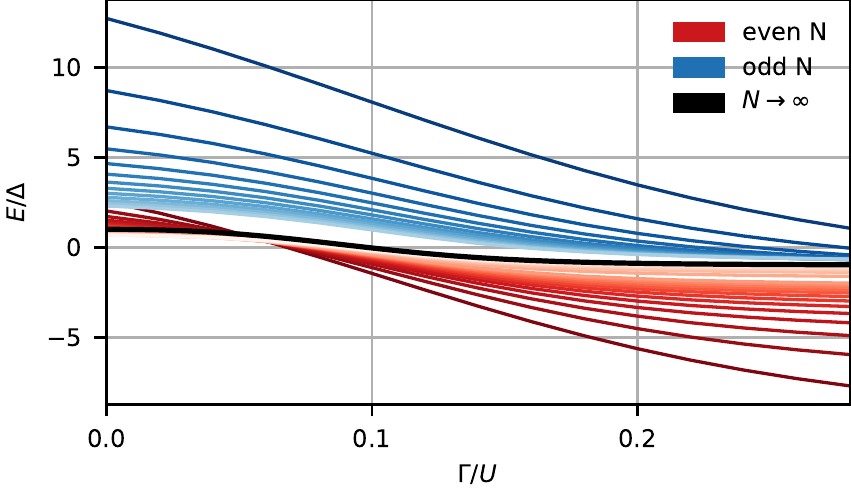}
\includegraphics[width=0.9\columnwidth]{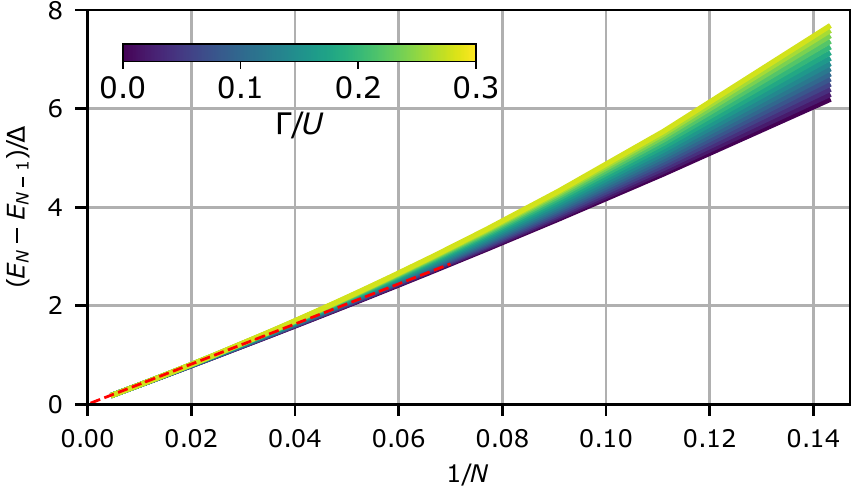}
\caption{
Finite-size effects in the YSR excitation spectra. Top: YSR energies normalized by the
superconducting gap at given $N$ (equal to the $\Gamma\to0$ limit of the subgap excitation energy). 
We plot all $N$ from 6 to 30, then pairs (40, 41), (50, 51), $\ldots$ (90, 91), then pairs (100, 101), (200, 201),
$\ldots$, (600, 601). Middle: YSR energies normalized by the BCS gap obtained in the large-$N$
limit. We plot all $N$ from 6 to 30. Bottom: $d \propto 1/N$ scaling of the difference between
even $N$ and odd $N$ results for a range of $\Gamma/U$ ratios. 
The asymptotic large-$N$ behavior of the difference is linear in $1/N$ (red dashed line).
Here $U=10$.
}
\label{N}
\end{figure}

Ultra-small superconducting islands have excitation spectra which
significantly differ from the BCS spectra in the large-$N$ limit, with
some elementary excitations which have no counterpart in the
Bogoliubov picture \cite{Roman2003,Yuzbashyan2003}. We assess the
effect of the ratio between the interlevel separation $d=2D/N$ and the
BCS gap $\Delta$ on the Yu-Shiba-Rusinov states of the DMRG solution
of the QD-SC problem in Fig.~\ref{N}. When scaled in terms of the
superconducting gap of the finite-size system (obtained as the
$\Gamma\to0$ limit of the YSR excitation energy), the curves tend to
approach the asymptotic YSR curve from above or below for odd and even
$N$, respectively. For $N=800$, used in most calculations in this
work, the Bogoliubov picture is valid and the results are even
quantitatively close to those for a superconductor in the
thermodynamic limit, although some finite-size corrections to the BCS
mean-field theory remain present. The superconducting islands in the
contemporary hybrid devices are sufficiently large that the effects
beyond the BCS theory need not be considered.

\subsection{Phase diagrams}

\begin{figure*}[htbp]
\centering
\includegraphics[width=1.8\columnwidth]{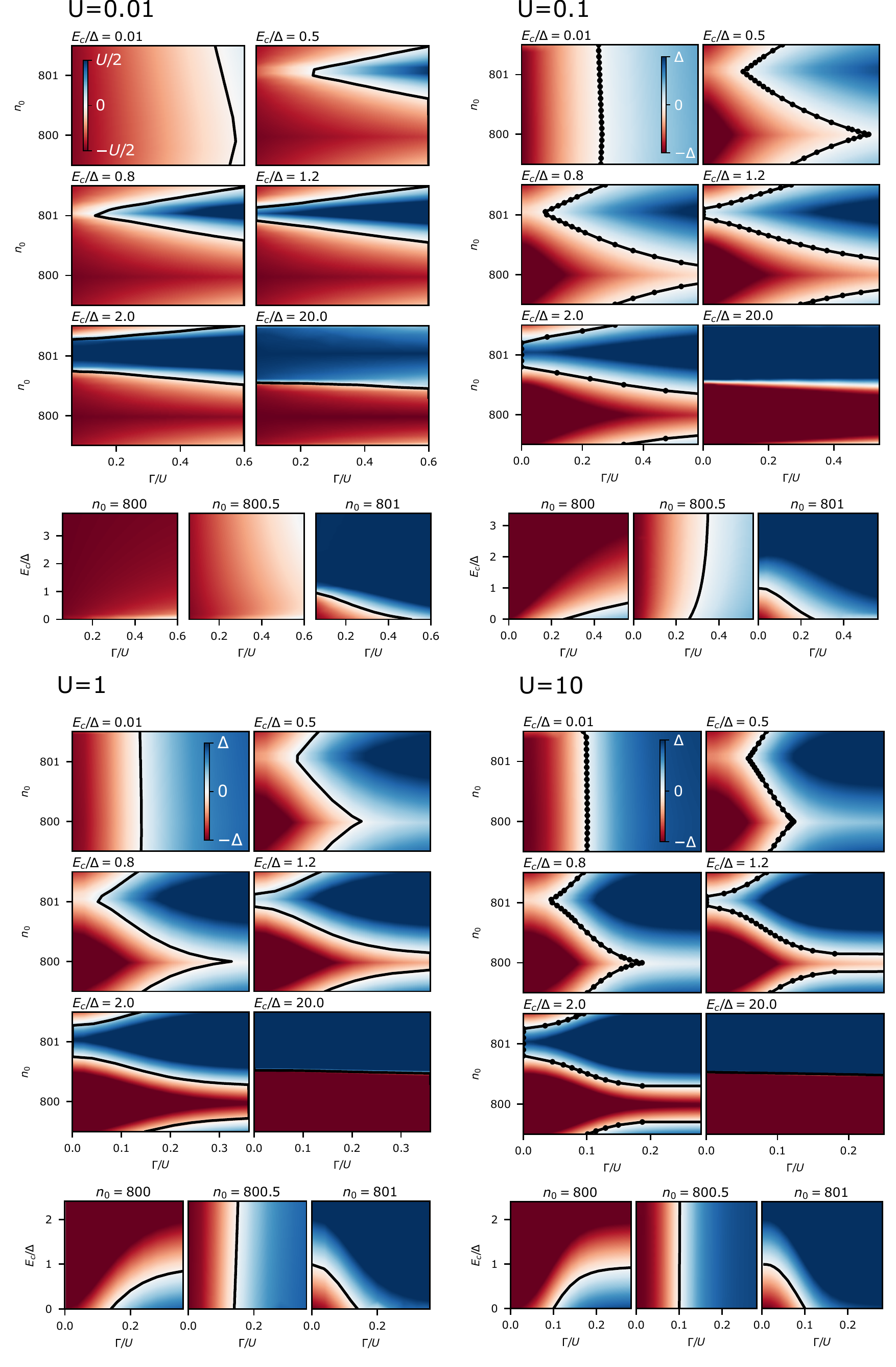}
\caption{Quantum phase transition between the doublet and singlet phases
for different values of $U$. The case of $U=0.1$ is shown in Fig.~1 of
the main text.}
\label{diag1bis}
\end{figure*}

Fig.~\ref{diag1bis} presents the phase diagrams at fixed impurity gate
voltage $\nu=1$ for several values of $U$ to supplement those for
$U=0.1\approx 4\Delta$ shown in Fig.~1 of the main text. 

For $U=10 \gg D,E_c,\Delta$ and $\nu=1$, the QD occupancy is pinned to
$n_\mathrm{imp} \approx 1$; the QD is then a pure exchange scatterer
with the Kondo exchange coupling constant $J_K=8\Gamma/\rho \pi U$. 
In this case, the phase diagram can be well reproduced using the following approximation to the many-body energy levels in
even/odd $\nsc$ sectors:
\begin{equation}
E(\nsc) \approx \text{const} + E_c(\nsc-n_0)^2 + \begin{cases}
0 & \nsc\text{ even,} \\
\Delta - E_B(\Gamma) & \nsc\text{ odd,}
\end{cases}
\label{Eq1}
\end{equation}
where $E_B(\Gamma)$ is the binding energy of the YSR quasiparticle,
such that $\lim_{\Gamma\to0} E_B(\Gamma) = 0$. The data for
$E_B(\Gamma)$ can be taken e.g. from a NRG calculation. This is the
idealized version of Fig.~1 from main text, showing qualitatively
similar evolution for YSR to CB regimes, but with large quantitative
differences, in particular in the cross-over $E_c \approx \Delta$
regime where the competition between various physical mechanisms is
the most pronounced. In the large-$U$ limit, at half-integer filling
of the superconductor ($n_0=800.5$) the transition between singlet and
doublet GS occurs at essentially the same value of $\Gamma$ for all
values of $E_c$ due to equal charging energies for both states, as
follows from Eq.~\eqref{Eq1}, because for large $U$ the only
dependence on $n_0$ is explicitly through the charging
term: the transition then occurs for $E_B(\Gamma)=\Delta$.

Eq.~\eqref{Eq1} also explains the change of topology in the phase
diagrams at $E_c = \Delta$. This occurs because the binding energy of
the YSR quasiparticles is bounded as $0 < E_B(\Gamma) <
2\Delta$. For $E_c>\Delta$ it is no longer possible to trap a
Bogoliubov quasiparticle at the impurity site for even $n_0$, thus the
system remains in a doublet ground state for any value of $\Gamma$.

For $U=1$, the system is essentially still in the deep
Kondo regime. The quantitative differences at $U=0.1$ (the value
corresponding to all results shown in the main text) are, however, significant.
They are most apparent in panel b) showing the phase diagram in
the $(\Gamma,E_c)$ plane. For even $n_0=800$, we observe a significantly slower approach to the
$E_c=\Delta$ asymptote compared to the large-$U$ limit. For half-integer $n_0=800.5$ the quantum
phase transition value of hybridisation ($\Gamma_C$) exhibits a weak $E_c$ dependence, thus
the transition line is no longer strictly vertical.  This is due to the competition between the
$E_c$ and $U$ terms, i.e., due to the redistribution of charge which is now possible even at the
half-way $n_0=800.5$ point because of the weaker electron-electron repulsion $U$ on the QD. The
diagram for odd $n_0=801$ appears to be less affected. 

Finally, for very low $U=0.01$ the system
is in a qualitatively different weak-interaction regime where the subgap states are better
described as {\sl Andreev bound states} (ABS). Here the increasing $E_c/\Delta$ ratio drives a
cross-over between the ABS and CB regimes that still shows some similarities with the YSR-CB
cross-over. The discussion of this regime is beyond the scope of this work.

\subsection{Charging diagrams}

\begin{figure*}[b]
\centering
\includegraphics[width=1.9\columnwidth]{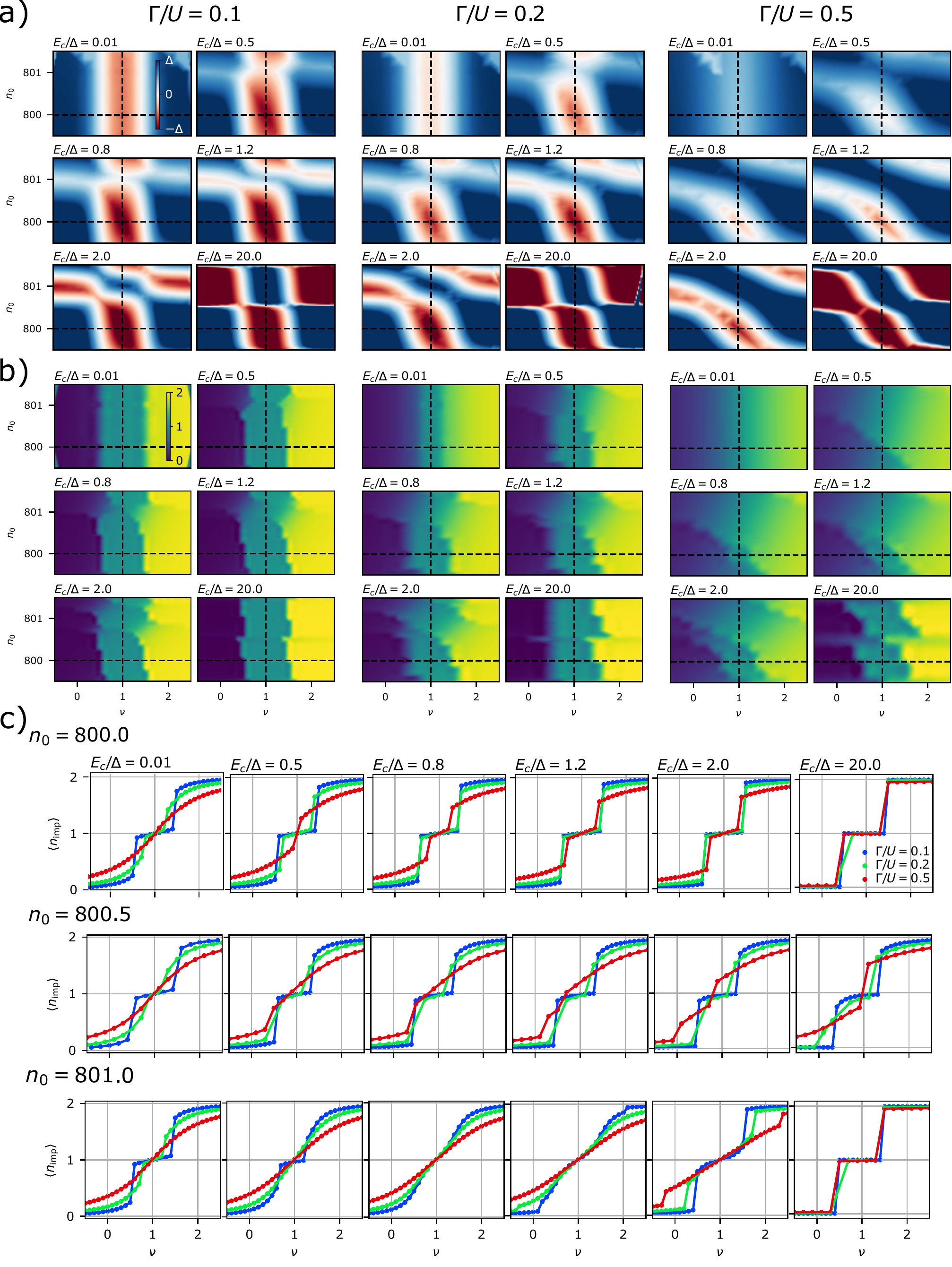}
\caption{a) Charging diagrams for a range of $\Gamma/U$. $\Gamma/U=0.1$ is the case shown in Fig.~2 of the main text.
b) Corresponding diagrams of the impurity occupancy, and c) cross-cuts at constant $n_0=800$,
$800.5$, and $801$.
}
\label{charging1}
\end{figure*}

In Fig.~\ref{charging1}a we show additional charging diagrams to supplement those shown in Fig.~2
of the main text. Increasing $\Gamma$ leads to more diffuse appearance of the charging patterns
and more dominant diagonal striping, which is a consequence of the formation of a "large single
quantum dot" comprising both the original QD and the SC island with the effective level controlled
by the sum of gate voltages, $n_0+\nu$. As an aid in the interpretation of these diagrams, in
Figs.~\ref{charging1}b and \ref{charging1}c we show the impurity occupancy using two representations:
as density plot that can be directly compared with the charging diagrams, and additionally as line
cuts at even $n_0=800$, half-integer $n_0=800.5$ and odd $n_0=801$. With increasing $\Gamma$ the
variation of $\langle n_\mathrm{imp} \rangle$ with $\nu$ becomes increasingly smooth and the local
magnetic moment for $\nu \approx 1$ becomes less defined (for $\Gamma/U=0.5$ one has $U/\pi \Gamma
\approx 0.64 < 1$, hence no local moment in the Hartree-Fock picture \cite{anderson1961}). This
effect is partly compensated by increasing $E_c$ which reduces the charge fluctuations betweed the
QD and the SC island. For instance, the local moment reemerges in the case of $\Gamma/U=0.5$ for
$E_c/\Delta \gtrsim 0.8$ when $n_0$ is even. This is concomitant with the appearance of a doublet
region in the phase diagram. For odd $n_0$ this process is less efficient, QD occupancy becomes
quantized only for very large values of $E_c$.

\subsection{Spectral functions}

\begin{turnpage}
\begin{figure*}[htbp]
\centering
\includegraphics[width=2.4\columnwidth]{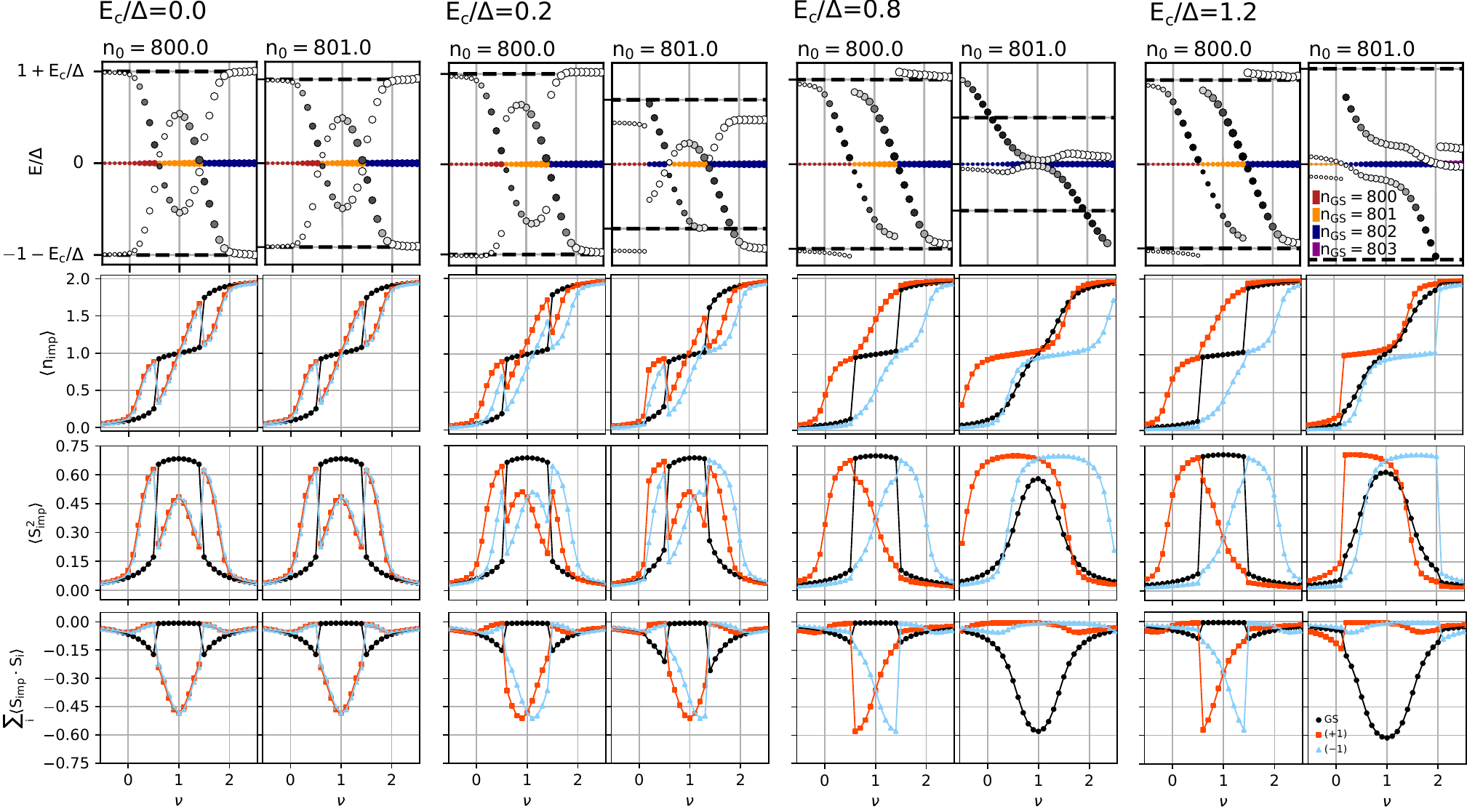}
\caption{Ground and excited state properties for even and odd integer
value of $n_0$. All parameters as in Fig.~3 of the main text,
in particular $U=0.1$, $\Gamma/U=0.1$.}
\label{Fig6}
\end{figure*}
\end{turnpage}

The subgap (discrete) part of the spectral function is easily computed from the wavefunctions
of (0), (+1) and (-1) states as 
\begin{equation}
\begin{split}
A_\sigma(\omega) &= | \langle \psi^{(+1)}_{\sigma'} | d^\dag_\sigma | \psi^{(0)}_{\sigma_0} \rangle|^2 \delta[\omega-
(E^{(+1)}-E^{(0)})] \\
&+ | \langle \psi^{(-1)}_{\sigma''} | d_\sigma | \psi^{(0)}_{\sigma_0} \rangle |^2
\delta[\omega+(E^{(-1)}-E^{(0)}].
\end{split}
\end{equation}
Here $\sigma'=\sigma_0+\sigma$ and $\sigma''=\sigma_0-\sigma$, where $\sigma_0$ is the $S_z$
component of total spin of the ground state (0), while $\sigma'$ and $\sigma''$ are those of the
excitations (+1) and (-1), respectively.

In Fig.~\ref{Fig6} we supplement the spectra shown in Fig.~3 of the main text with the additional
results that reveal the nature of the relevant ground and excited states, and show the evolution
with increasing $E_c$ leading to the $E_c > \Delta$ regime (Fig.~4 of the main text).
Specifically, in Fig.~\ref{Fig6} we compare the $\nu$ dependence of the subgap state energies,
spectral weights, and the expectation values of occupancy, local moment, and spin-spin
correlation for the range of $E_c$ discussed in the main text.

The left-most panels show the reference results for $E_c=0$, i.e. the conventional YSR regime. The
model parameters used here correspond to the situation where close to half filling ($\nu=1$,
particle-hole symmetric point) the ground state is an (unscreened) doublet, while sufficiently
away from half filling it is a singlet. At $\nu \sim 1$, the doublet GS with $\langle n_\mathrm{imp} \rangle
\approx 1$ is characterized by a nearly saturated local moment $\langle \mathbf{S}^2_\mathrm{imp}
\rangle \to 3/4$ that is almost decoupled from the band electrons, $\sum_i \langle
\mathbf{S}_\mathrm{imp} \cdot \mathbf{S}_i \rangle \approx 0$.
Sufficiently away from $\nu=1$, the singlet GS has level filling
closer to either zero or full (double) occupancy and correspondingly diminished local moment $\langle
\mathbf{S}^2_\mathrm{imp} \rangle \lesssim 0.15$. The excited states in the central $\nu\sim1$ region have
level filling with a strong dispersion, and there is strong antiferromagnetic alignment
of the local moment with the electrons
in the band: this is a manifestation of the bonding of the Bogoliubov quasiparticle that generates
these subgap YSR states. We note some small differences in the results for (+1) and (-1) excited
states (in particular the red and blue symbols do not overlap completely): this is a consequence
of the finite size (finite $N$) of the system, as discussed earlier. For $E_c=0$ the results do not depend on $n_0$. The
results are therefore (anti)symmetric with respect to $\nu=1$ for both even and odd $n_0$.

For finite but small $E_c=0.2\Delta$, the main qualitative difference compared to $E_c=0$ is the
observably different behavior of the (+1) and (-1) excitations, far exceeding the finite-$N$ effects
we noted for the case of $E_c=0$. Furthermore, we observe a lack of (anti)symmetry with respect
to $\nu=1$ for odd $n_0=801$. The nature of the states remains, however, the same as for $E_c=0$.

The regime of $E_c/\Delta=0.8$ and $E_c/\Delta=1.2$, where the electron-electron repulsion terms 
on the superconductor, $E_c$, and on the quantum dot, $U$, are comparable in magnitude
(specifically $E_c/(U/2)\approx0.4,0.6$), is controlled by the competition between the QD and SC
filling. In the following we analyse this regime in more details, separately for even
and odd $n_0$.

For even $n_0$, close to the p-h symmetric point ($\nu\approx1$), the ground state (0) is a
doublet with the impurity local moment almost decoupled from the SC. The excited states ($\pm1$)
are similar to conventional YSR singlets, but with $n_\mathrm{imp}$ considerably different from 1
due to charging terms.

Away from half filling, for large values of $\nu$ ($\nu \gtrsim 1.5$), in the state (0) there are
almost 2 electrons on the impurity and $\nsc$ is even. In (+1), the additional electron enters an
empty SC level, which costs $E_c+\Delta$, thus $E^+$ lies at the bottom of the continuum.  To
obtain (-1), the electron removed from the GS does not originate entirely from the SC but also
partly from the impurity ($\langle n_\mathrm{imp} \rangle < 2$ in the hole-like excited state),
recovering some of the local moment $\langle S^2_\mathrm{imp} \rangle$ and shifting $E^-$ inside
the gap due to hybridisation.

For odd $n_0$, close to the p-h symmetric point the GS (0) and the excited states $(\pm1)$ all
have an occupancy close to 1 and a well developed local moment. The difference between these states
consists in the fate of this moment: in the GS it forms a strong QD-SC singlet state with the lone
Bogoliubov quasiparticle, while in $(\pm 1)$ states it is simply decoupled. The excitation
energies are given by the sum of $-\Delta+E_c=0.2\Delta$ and a contribution proportional to $J_K$,
as discussed in the main text.

For large values of $\nu \sim 2$, the competition between $E_c$ and $U$ is very prominent and we
need to distinguish the regions where the GS has 802 or 803 electrons; the transition between them
occurs at a value of $\nu$ that strongly depends on $E_c/\Delta$. In the region with
$n_\mathrm{GS}=802$, the state (0) is a singlet with high $n_\mathrm{imp}$. It has a large SC
charging energy, while the impurity e-e repulsion energy $\frac U2 (n_\mathrm{imp}-\nu)^2$ is
almost minimized. An additional electron predominantly enters the SC, hence the peak $E^+$ has low
spectral weight. The states (-1) and (0) are, however, quite similar, except for the additional
electron at the impurity site in the state (0). The spectral weight of the $E^-$ peak is thus
large, on the order of $0.85$, and the corresponding excitation energy is large because the
low-energy impurity level is emptied. In the region with $n_\mathrm{GS}=803$, the states (0), (-1)
and (+1) all have close to maximal $n_\mathrm{imp}$, and differ only in the electron close to the
Fermi level in the SC, leading to extremely small spectral weights of the subgap peaks.

\begin{figure}
\centering
\includegraphics[width=0.9\columnwidth]{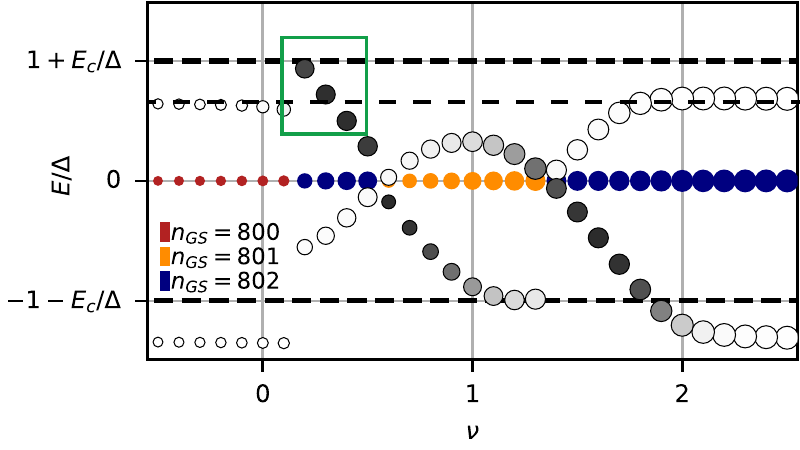}
\caption{Spectral function for odd $n_0=801$ and $E_c/\Delta=0.2$, the case shown in Fig.~3f) in
the main text. The thin dashed line at $\Delta-E_c$ shows the edge of the continuum for the
isolated SC island ($\Gamma=0$ limit).
}
\label{onkraj}
\end{figure}

We note the existence of cases where the ``subgap'' state energy exceeds the $\Gamma=0$ gap edge,
see Fig.~\ref{onkraj}. At $E_c/\Delta=0.2$ with odd $n_0$, the gap edge for $E_c<\Delta$ (thin
dashed line) is given by $\Delta-E_c$ (see section~\ref{edges} below for derivation). In the bias
voltage range indicated by the box in Fig.~\ref{onkraj}, the ``subgap'' state crosses this line
{\sl with a finite weight} of the spectral peak, in striking contrast to the usual situation where
the YSR peaks transfer weight continuously as they approach the gap edge when the bound state
merges with the continuum. This effect occurs away from $\nu=1$ for finite $E_c$ and odd $n_0$, in
situations where it is advantageous for the tunneling electron to occupy the impurity orbital
rather than enter a SC level.

\subsection{Extraction of $E_c$ from discontinuities}

\begin{figure}
\centering
\includegraphics[width=0.9\columnwidth]{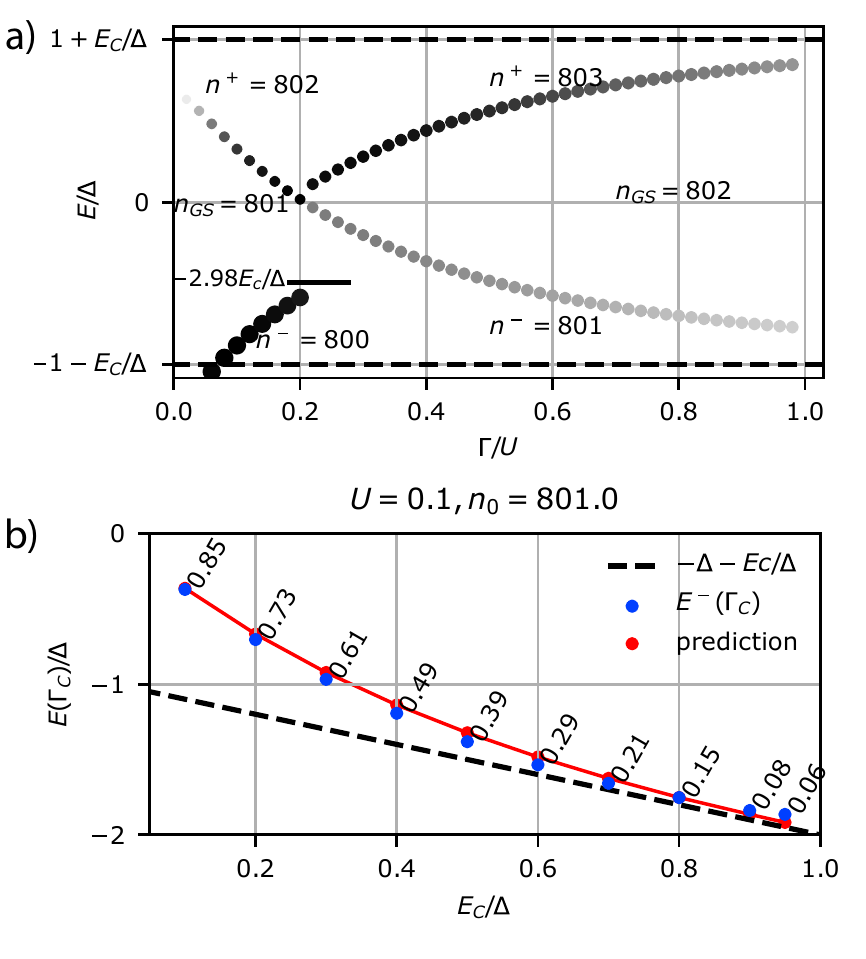}
\caption{Quantifying $E_c$ from the discontinuities in spectral functions at odd SC filling, $n_0=801$.
Top: spectral function as a function of hybridisation $\Gamma$,
across the doublet-singlet transition. Here $U=0.1$, $E_c/\Delta=0.2$, $\nu=1$ (the case of Fig.~\ref{fig4}f).
Bottom: discontinuity $E^-$ as a function of $E_c$ (blue dots). Red line indicates an estimated based on electrostatic energy (see text)
taking into account the impurity occupancy at the transition point (black labels). Dashed line indicates the gap edge.
$U=0.1$, $\nu=1$.
}
\label{fig5}
\end{figure}

The p-h asymmetry and discontinuities provide a means to determine the charging energy $E_c$ from
experimental spectra. This is best done for the system tuned to odd $n_0$ where the asymmetries
are maximal. In Fig.~\ref{fig5}(a) we plot the $\Gamma$-dependence of the peak positions for an
impurity tuned to $\nu=1$.  For $E_c=0$, $E^{(0)}$, $E^{(+1)}$ and
$E^{(-1)}$ would all be equal at the doublet-single transition point
$\Gamma=\Gamma_c$. For $E_c>0$, this no longer holds for odd $n_0$, as $\nsc$ of (-1) differs from
$n_0$. The asymmetry $E^+-E^-$ is proportional to $E_c$, with a prefactor that depends on the
impurity occupancy and in general needs to be determined numerically, see Fig.~\ref{fig5}(b). In the
large-$U$ (Kondo) limit where $n_\mathrm{imp}\approx 1$, the energy difference is simply $4E_c$
and $E_c$ could be directly extracted from experimental measurements. For comparable values of $E_c$
and $U$, this is no longer the case and the competition between the QD and SC charging terms is
observed. In this case, $E^+-E^-$ can be approximated by the difference of the sums of the
impurity and SC charging energies in each state: $E^+ - E^- =
\Big(
E_\mathrm{imp}^{(+1)} + E_c\big(n^{(+1)}_{sc}-n_0\big)^2
\Big)
-
\Big(
E_\mathrm{imp}^{(-1)} + E_c\big(n^{(-1)}_{sc}-n_0\big)^2
\Big)$, where $E^x_\mathrm{imp}$ is the expectation value of $H_\mathrm{imp}$ in the sector
indicated by the superscript label $x \in \{ (+0),(-1),(+1) \}$. This estimate is plotted in Fig.~\ref{fig5}
as the red line labelled ``prediction''. The good agreement with the exact results indicates that
these effects are indeed controlled mainly by the charging terms, while the hybridisation energy
is roughly the same in $(+1)$ and $(-1)$. In order to experimentally determine $E_c$ in this regime,
it is necessary to have either accurate information about the impurity occupancy, or make use of
numerical calculations to fit the experimental results and extract the model parameters.

\section{Model definition}

Here we provide some further details on the Hamiltonian used in the
main text.
We first introduce the notation for level and electron numbers. $N$ is the number of levels
in the SC, $M=N+1$ is the total number of levels in the problem.
We furthermore define the following occupancy operators:
\begin{equation}
\hat{n}_{\mathrm{imp},\sigma} = d^\dag_\sigma d_\sigma,\quad
\hat{n}_\mathrm{imp} = \hat{n}_{\mathrm{imp},\uparrow} + \hat{n}_{\mathrm{imp},\downarrow},
\end{equation}
for the impurity, and
\begin{equation}
\hat{n}_\mathrm{sc}=\sum_{i\sigma} c^\dag_{i\sigma} c_{i\sigma}
\end{equation}
for the SC, as well as $\hat{n} = \hat{n}_\mathrm{imp} + \hat{n}_\mathrm{sc}$.
We write $n_\mathrm{imp} = \langle \hat{n}_\mathrm{imp} \rangle$,
$n_\mathrm{sc} = \langle \hat{n}_\mathrm{sc} \rangle$ and $n = \langle \hat{n} \rangle$.
Evidently, $n_\mathrm{imp} + n_\mathrm{sc} = n$.
At half filling $n_\mathrm{imp}=1$, $n_\mathrm{sc}=N$, and $n=1+N=M$.

The SC parts of the Hamiltonian are
\begin{equation}
\begin{split}
H'_\mathrm{sc} &= \sum_{i,\sigma} \epsilon_i c^\dag_{i\sigma} c_{i\sigma}
- \alpha d \sum_{i,j}^N c^\dag_{i\uparrow} c^\dag_{i\downarrow}
c_{j\downarrow} c_{j\uparrow},
\\
H''_\mathrm{sc} &= E_c (\hat{n}_\mathrm{sc} - n_0)^2.
\end{split}
\end{equation}
The pairing terms only include the time-reversal conjugate states, i.e., the
Hamiltonian takes the form of the reduced pairing model (see also Appendix C in Ref.~\onlinecite{Braun1999}).
$\alpha$ is the (dimensionless) strength of the attractive electron-electron interaction.
$E_c$ is the charging energy of the SC island, $E_c = e_0^2/2C$, where $C$ is the total
capacitance of the island. The interlevel spacing is $d=2D/N$, where $2D$
is the bandwidth of the conduction band. More precisely,
the energy levels are $\epsilon_i = -D + d/2 +(i-1)d + x$ for $i=1,\ldots,N$, so that
$\epsilon_1=-D+d/2+x$ and $\epsilon_N = +D-d/2+x$. Here $x=\alpha d/2$ is a small shift that restores
the p-h symmetry of the finite-size problem, as discussed in a later subsection.

The impurity part of the Hamiltonian is
\begin{equation}
H_\mathrm{imp} = \sum_\sigma \epsilon_\mathrm{imp} \hat{n}_{\mathrm{imp},\sigma}
+U \hat{n}_{\mathrm{imp},\uparrow} \hat{n}_{\mathrm{imp},\downarrow}.
\end{equation}
We introduce $\delta=\epsilon_\mathrm{imp}+U/2$, as well as
$\nu = 1/2 - \epsilon/U = 1 - \delta/U$,
as two further ways to express the value of $\epsilon_\mathrm{imp}$ for a given fixed
value of $U$. Both measure the departure from the p-h symmetric point at
$\delta=0$ and $\nu=1$, $\delta$ in energy units, $\nu$ in units of electron number.
Thus, alternatively,
\begin{equation}
\begin{split}
H_\mathrm{imp} &= \frac{U}{2} (\nimp-1)^2 + \delta (\nimp-1) + \text{const} \\
&= \frac{U}{2} (\nimp-\nu)^2 + \text{const}.
\end{split}
\end{equation}

Finally, the hybridisation part is
\begin{equation}
H_\mathrm{hyb} = \frac{v}{\sqrt{N}} \sum_{i,\sigma}^N \left(
c^\dag_{i\sigma} d_\sigma + \text{h.c.} \right).
\end{equation}
We define $\Gamma=\pi \rho v^2$ where $\rho=1/2D$ is the density of states in
the conduction band. Conversely, $v=\sqrt{\Gamma/\pi \rho} = \sqrt{2\Gamma/\pi}$.

In a strictly electrically isolated QD-SC system the number of electrons would be fixed.
The presence of weakly coupled tunneling probes permits the transfer of charge to and from
the QD-SC system. The total occupancy changes so as to reach the state of minimal energy.
(Strictly speaking the thermodynamic variable that is minimized is $H-\mu n$, but $\mu$ may be
thought to be absorbed in the parameters $n_0$ and $\nu$).
Nevertheless, for weak tunneling probe coupling the charge fluctuations may be neglected
and the total QD-SC system charge does not change with time beyond the tunneling
events when the system is probed. For this reason, one may take $\langle \hat{n} \rangle \equiv n$
to be an integer constant.
Of particular interest is the parity of $n$ in the ground state. In the usual discussions of the YSR
physics, where $E_c=0$ and $n_\mathrm{imp} \approx 1$, the odd-parity state (doublet state)
corresponds to an unscreened impurity, the even-parity state (singlet state) to a ``YSR screened''
impurity. At finite $E_c$ this picture is modified by the additional energy shift of $E_c$ for
states with SC occupancy differing by one electron.

We denote the lowest-energy eigenstate in each charge sector as $\psi^n$ and its energy as $E^n$.
The ground state is thus $\psi^{\ngs}$ and the lowest excited states $\psi^{\ngs+1}$ and
$\psi^{\ngs-1}$. The excitation energies of the spectroscopically visible subgap states are
defined as $E^+=E^{(+1)}-E^{(0)}$ for particle addition and $E^- = E^{(-1)}-E^{(0)}$ for particle
removal. The corresponding spectral weights are $w^+ = |\langle \psi^{(+1)} | d^\dag_\sigma |
\psi^{(0)} \rangle|^2$ and $w^- = |\langle \psi^{(-1)} | d_\sigma | \psi^{(0)} \rangle|^2$.

It is perhaps worthwhile to point out that since our model is based on a ``microscopic'' description of
the pairing interaction and since the Hamiltonian is solved essentially exactly within the DMRG,
the ``gap renormalization effects'' (the effect of the impurity back on the superconductor)
is fully taken into account, thus no self-consistent correction of the pairing function
is necessary as in mean-field approaches. This $1/N$ renormalization effect is, however,
small even for the $N=800$ SC levels used in the calculations in this work.

\subsection{Continuum edges}
\label{edges}

For easy reference, let us consider the lowest particle-addition and particle-removal excitation
energies for a pure SC island in the absence of the QD, i.e., the edges of the quasi-continua
of Bogoliubov quasiparticles for finite $E_c$.

From Eq.~\eqref{Eq1} (for $\Gamma \equiv 0$), we find that for a GS with an even integer
occupancy of the superconductor $n_\mathrm{sc}$
\begin{equation}
\begin{split}
E^+ &= \Delta+E_c+2E_c(\nsc-n_0), \\
E^- &= \Delta+E_c-2E_c(\nsc-n_0).
\end{split}
\end{equation}
Here $n_0$ is the continuously tunable experimental parameter proportional to gate voltage, 
while $n_\mathrm{sc} = \langle \hat{n}_\mathrm{sc} \rangle$ is an integer (except at the charge degeneracy points).
This reduces to $E^+=E^-=\Delta+E_c$ for even integer $n_0=\nsc$, but one should note that $E^+$
and $E^-$ are shifted {\it asymmetrically} for any value of $n_0$ that is not an even integer. The
total single-particle gap in the spectral function thus remains constant, $E^+ + E^- = 2\Delta +
2E_c$. The largest asymmetry occurs for values close to odd $n_0$. Exactly at odd-integer $n_0$,
$\nsc$ changes discontinuously by 2 for $E_c < \Delta$. On one side of this discontinuity one finds
\begin{equation}
\begin{split}
E^+ &= \Delta-E_c, \\
E^- &= \Delta+3E_c,
\end{split}
\end{equation}
and on the other
\begin{equation}
\begin{split}
E^+ &= \Delta+3E_c, \\
E^- &= \Delta-E_c.
\end{split}
\end{equation}
For $E_c=0$ we recover the standard BCS result with the SC gap edges at $\omega=E^+=\Delta$
and $\omega=-E^-=-\Delta$ for all values of $n_0$.

For large $E_c > \Delta$, the excitation gaps of the even $\nsc$ states close at
\begin{equation}
\begin{split}
n_0 &= \nsc + \frac{1}{2}(1+\Delta/E_c), \\
n_0 &= \nsc - \frac{1}{2}(1+\Delta/E_c),
\end{split}
\end{equation}
for particle-addition and particle-removal gap, respectively. In the range of $n_0$ where the GS
has an odd number of electrons in the SC the following expressions hold:
\begin{equation}
\begin{split}
E^+ &= -\Delta+E_c+2E_c(\nsc-n_0), \\
E^- &= -\Delta+E_c-2E_c(\nsc-n_0).
\end{split}
\end{equation}
For odd $\nsc=n_0$ we thus find $E^+=E^-=-\Delta+E_c$.

\begin{figure*}[htbp]
\centering
\includegraphics[width=0.5\columnwidth]{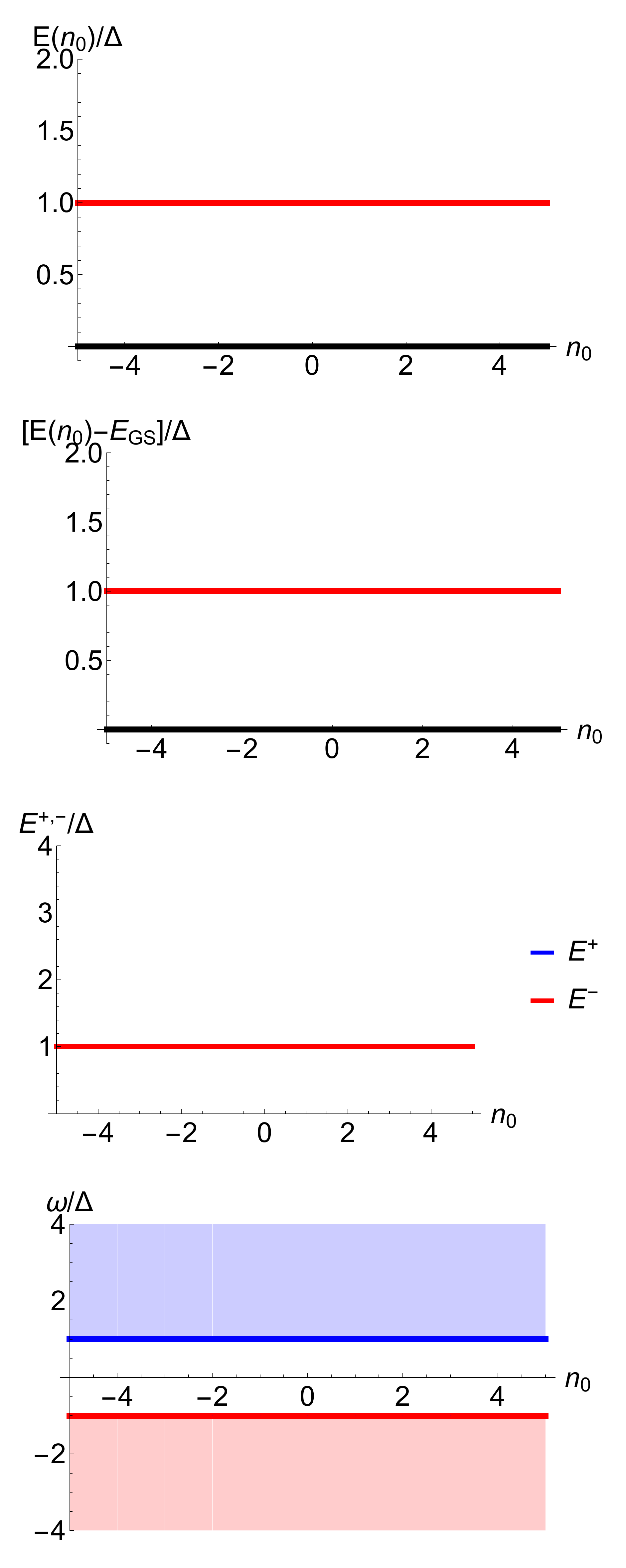}
\includegraphics[width=0.5\columnwidth]{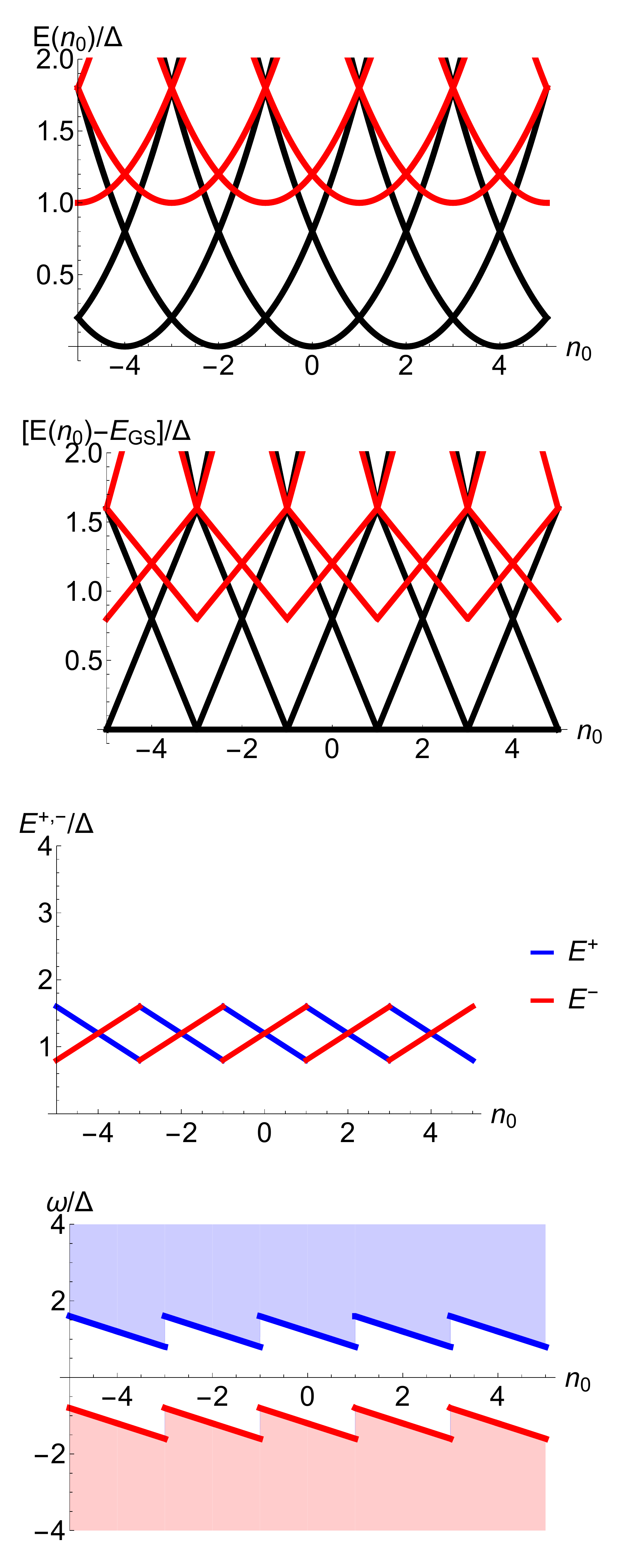}
\includegraphics[width=0.5\columnwidth]{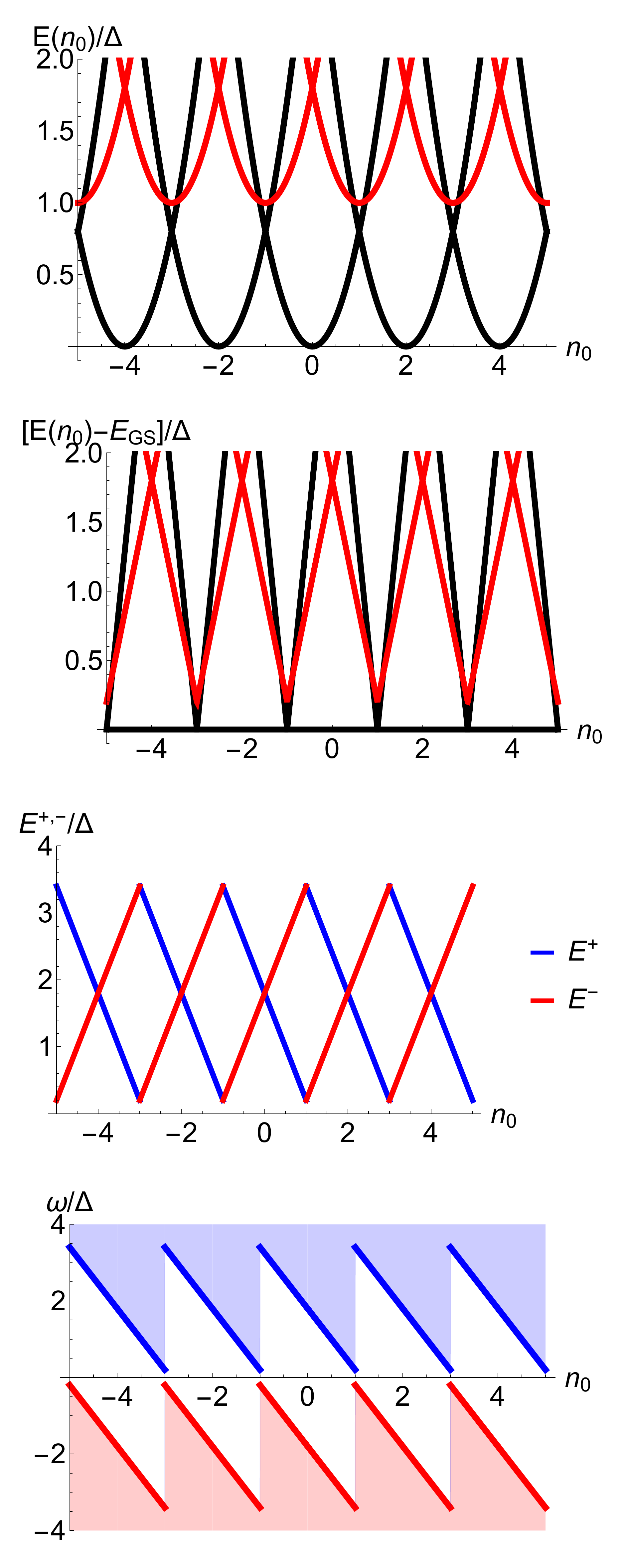}
\includegraphics[width=0.5\columnwidth]{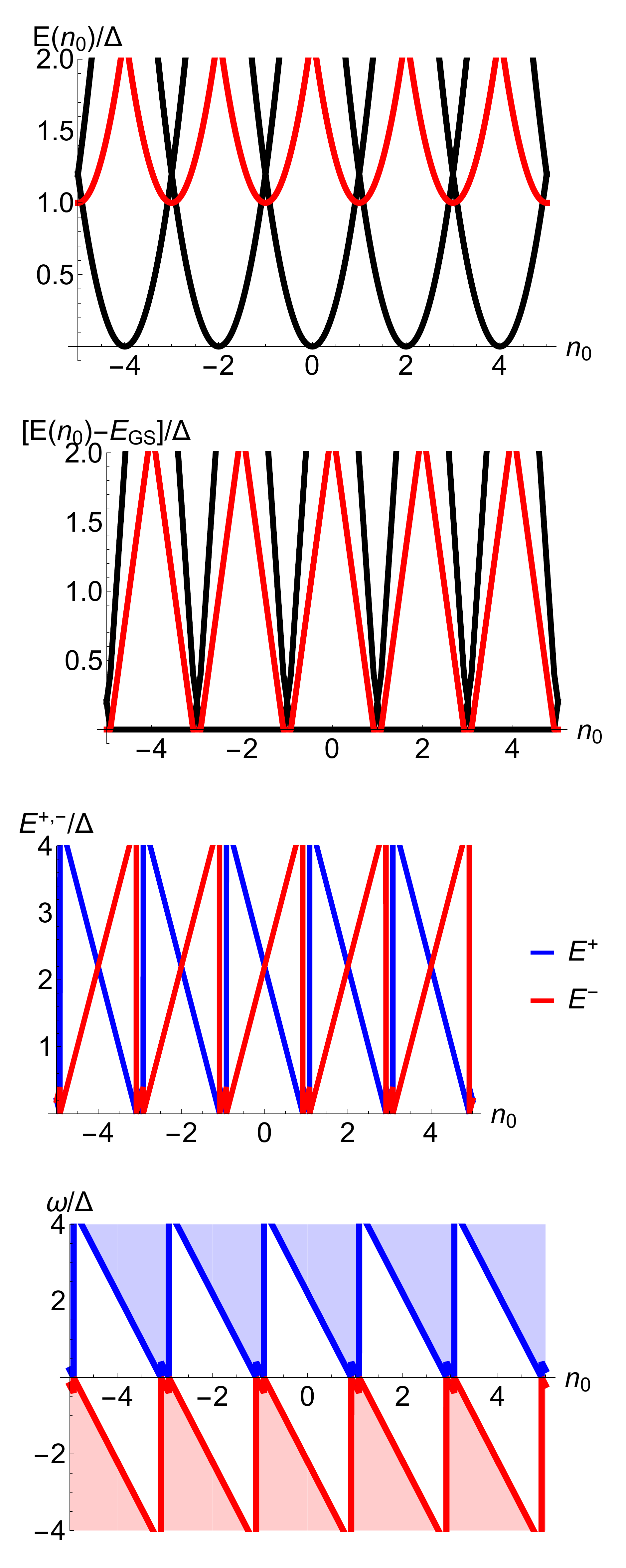}
\caption{Single-particle excitation properties of the SC island in the absence of the QD, for
(left to right) $E_c/\Delta=0, 0.2, 0.8, 1.2$. In top rows, black stands for even $\nsc$, red for
odd $\nsc$. For non-zero $E_c$, the charging has period $2e$. For $E_c/\Delta<1$, the occupancy
changes in steps of 2. For $E_c/\Delta>1$, the occupancy changes in steps of 1, yet the regions of
even and odd occupancy have different widths and the $2e$ period is maintained. With increasing
$E_c/\Delta$, the even-odd effects become less pronounced; in the large $E_c/\Delta$ limit a
$1e$-periodic pattern typical of Coulomb blockade is recovered. At the same time, with increasing
$E_c/\Delta$ the nature of the gap is changing from the superconducting gap into a {\sl Coulomb
gap}.}
\label{overview}
\end{figure*}

In Fig.~\ref{overview} we show a graphical overview of these results. For $E_c/\Delta=0, 0.2, 0.8, 1.2$ (the
values used in Fig.~3 of the main text) we plot in separate rows i) the energies of the SC states,
ii) these same energies referred to the lowest (GS) energy, iii) the excitation energies $E^+$ and $E^-$, iv) the edges of
the continuum of single-particle (Bogoliubov) excitations at $\omega=+E^+$ and $\omega=-E^-$.

\subsection{Particle-hole symmetry}

The particle-hole (p-h) transformation is defined as
\begin{equation}
\begin{split}
d^\dag   &\to d, \\
c^\dag_i &\to -c_{N+1-i}.
\end{split}
\end{equation}
The Hubbard and hopping terms remain invariant. The charge terms
transform as $\nimp \to 1-\nimp$, $c^\dag_{i,\sigma} c_{i,\sigma}
\to 1-c^\dag_{N+1-i,\sigma} c_{N+1-i,\sigma}$, so that
$\nsc \to N-\nsc$. Finally, the pairing terms transform
as $\sum_{i,j} c^\dag_{i\uparrow} c^\dag_{i\downarrow} c_{j\downarrow} c_{j\uparrow}
\to N - \sum_i \sum_\sigma c^\dag_{i\sigma} c_{i\sigma} + \sum_{i,j} c^\dag_{i\uparrow} c^\dag_{i\downarrow}
c_{j\downarrow} c_{j\uparrow}$.
The Hamiltonian thus remains invariant if
\begin{equation}
\begin{split}
\epsilon_\mathrm{imp} &= -U/2, \\
\epsilon_i &= -\epsilon_{N+1-i} + g.
\end{split}
\end{equation}
The solution to the second equation for equidistant levels with spacing $d$
is
\begin{equation}
\begin{split}
\epsilon_i &= -D + \frac{d}{2} + (i-1)d + \frac{\alpha d}{2} \\
&= -D + \left( i - \frac{1-\alpha}{2} \right) d.
\end{split}
\end{equation}
In the $N\to\infty$ limit, this converges to a flat band
with the density of states $\rho=1/2D$ on the interval $[-D:D]$.

\subsection{Truncation of the SC levels}

In Hamiltonian $H'_\mathrm{sc}$, we truncate the spectrum of the SC levels at the Debye frequency $\omega_D$.
For $\Gamma=0$ this is no approximation, because the levels outside the range $[-\omega_D:\omega_D]$
play no role since they are fully decoupled from the levels participating in pairing
\cite{Richardson1963, Richardson1964, Richardson1966, vonDelft1999, Sierra2000}.
For $\Gamma \neq 0$, equating $D=\omega_D$ is an approximation, since the SC levels
in ranges $[-D:-\omega_D]$ and $[\omega_D:D]$ are omitted. If required, one could explicitly take into
account these non-interacting levels through the renormalization of model parameters \cite{Hewson} using,
for example, the numerical renormalization group (NRG) method \cite{wilson1975,bulla2008}.
An alternative correction scheme is to consider the cut-off $\omega_D$ to be increased to $D$, while
the coupling constant $\alpha$ is decreased accordingly so that the gap $\Delta$, estimated through
the BCS relation $\Delta=\omega_D \exp(-1/\alpha d)$ remains constant.
In any case, the approximation $\omega_D=D$ has no qualitative effect on the results.

Another observation is that typically the bath has a very large number
of levels, while only a tiny subset of those is actually hybridized
with the impurity. The effective quantum impurity problems with
noninteracting baths fully disregard all levels which are decoupled
from the impurity, because those live in a separate Hilbert space and
are irrelevant for the solution of the impurity problem. In our work
we also retain only the levels that hybridise with the impurity and
neglect all others, but we need to keep in mind that in reality the
Coulomb interaction connects the two subsystems which therefore do not
fully decouple. We do not discuss effects resulting from such coupling
in this work, but merely note that they are expected to be important
for the transport properties.

\begin{widetext}
\subsection{Implementation of the method}

We first provide the matrix-product-operator (MPO) representation of the Hamiltonian studied in
this work. Left-most site (impurity-site):
\begin{equation}
W_0 = \begin{pmatrix}
I &
\epsilon_\mathrm{imp} \hat{n}_\mathrm{imp} + U \hat{n}_{\mathrm{imp},\uparrow}\hat{n}_{\mathrm{imp},\downarrow} &
-d_{\uparrow}F &
-d_{\downarrow}F &
+d^\dag_{\uparrow}F &
+d^\dag_{\downarrow}F &
0 &
0 &
0
\end{pmatrix}.
\end{equation}
Here $F=(-1)^n$ is the local fermionic-parity operator, which gives phase of $-1$ if there is an odd number of
electrons on the site.

Generic site (with $g=\alpha d$):
\begin{equation}
W_i = \begin{pmatrix}
1 & [\epsilon_i + E_c(1-2n_0)] \hat{n}_i + (g+2E_c) \hat{n}_{i\uparrow} \hat{n}_{i\downarrow} & 0 & 0 & 0 & 0 &
g c_{i\downarrow} c_{i\uparrow} & g c^\dag_{i\uparrow} c^\dag_{i\downarrow} &
2 E_c \hat{n}_{i} \\
0 & I & 0 & 0 & 0 & 0 & 0 & 0 & 0 \\
0 & v c^\dag_{i\uparrow} & F_i & 0 & 0 & 0 & 0 & 0 & 0 \\
0 & v c^\dag_{i\downarrow} & 0 & F_i & 0 & 0 & 0 & 0 & 0 \\
0 & v c_{i\uparrow} & 0 & 0 & F_i & 0 & 0 & 0 & 0 \\
0 & v c_{i\downarrow} & 0 & 0 & 0 & F_i & 0 & 0 & 0 \\
0 & c^\dag_{i\uparrow} c^\dag_{i\downarrow} & 0 & 0 & 0 & 0 & I & 0 & 0 \\
0 & c_{i\downarrow} c_{i\uparrow} & 0 & 0 & 0 & 0 & 0 & I & 0 \\
0 & \hat{n}_i & 0 & 0 & 0 & 0 & 0 & 0 & I
\end{pmatrix},
\end{equation}
with
\begin{equation}
\hat{n}_{i\sigma} = c^\dag_{i\sigma} c_{i\sigma}, \quad
\hat{n}_i = \sum_\sigma \hat{n}_{i\sigma},
\end{equation}
and $F_i$ is again a local parity operator.

Right-most site:
\begin{equation}
W_N = \begin{pmatrix}
[\epsilon_N + E_c(1-2n_0)] \hat{n}_N + (g+2E_c) \hat{n}_{N\uparrow} \hat{n}_{N\downarrow} \\
I \\
v c^\dag_{N\uparrow} \\
v c^\dag_{N\downarrow} \\
v c_{N\uparrow} \\
v c_{N\downarrow} \\
c^\dag_{N\uparrow} c^\dag_{N\downarrow} \\
c_{N\downarrow} c_{N\uparrow} \\
\hat{n}_N
\end{pmatrix}.
\end{equation}

An alternative representation is possible where the impurity is located in the center of the 1D
chain (corresponding to the Fermi level of the superconducting island) rather than attached to the
end of the chain. We find fully equivalent results with both approaches, with rather similar bond
dimensions (which are maximal in the vicinity of the Fermi level), similar to what has been
observed in solving impurity models in the star geometry using the DMRG method \cite{Wolf2014}.

The DMRG calculations have been performed using the ITensor library. The initial state is the
Fermi sea with all low-lying levels of the SC occupied by electrons, and an additional electron on
the impurity site. A low truncation criterion (sum of discarded Schmidt values)
$\epsilon=10^{-12}$ and bond dimensions up to 5000 are required in order to reach convergence for $N=800$.

The symmetries exploited in the DMRG calculations were the charge conservation $U(1)$ and the
spin conservation $U(1)$. The calculations were performed for $S_z=0$ in even-occupancy sectors
and for $S_z=\pm 1/2$ in odd-occupancy sectors.

We note that our method is very different from that in Ref.~\onlinecite{Dukelsky1999} which is a momentum-space DMRG in the space of
electron pairs (``particle-hole'' method), and hence inapplicable to our Hamiltonian that explicitly breaks electron pairs through exchange processes.

\end{widetext}

\section{Simplification to $8 \times 8$-dimensional MPOs}

We now consider the parts of the Hamiltonian which control the occupancy,
specifically:
\begin{equation}
H' = U \hat{n}_{\mathrm{imp},\uparrow} \hat{n}_{\mathrm{imp},\downarrow}
+\epsilon \hat{n}_\mathrm{imp} + E_c (\hat{n}_\mathrm{sc} - n_0)^2.
\end{equation}
We note that
\begin{equation}
\begin{split}
(\hat{n}_\mathrm{sc} - n_0)^2
&= \left[ \left(\hat{n}-\hat{n}_\mathrm{imp}\right)-n_0 \right]^2 \\
&= \left[ \left(\hat{n}-n_0\right)-\hat{n}_\mathrm{imp} \right]^2 \\
&= (\hat{n}-n_0)^2 -2(\hat{n}-n_0) \hat{n}_\mathrm{imp} + \hat{n}_\mathrm{imp}^2
\end{split}
\end{equation}
and
\begin{equation}
\begin{split}
\nimp^2 &= (\niu+\nid)^2 \\
&= \niu^2 + \nid^2 + 2\niu \nid \\
&= \niu + \nid + 2 \niu\nid \\
&= \nimp + 2\niu\nid,
\end{split}
\end{equation}
so that
\begin{equation}
(\nsc-n_0)^2 = (\hat{n}-n_0)^2 + [1-2(\hat{n}-n_0)] \nimp + 2 \niu\nid.
\end{equation}
Thus
\begin{equation}
\begin{split}
H' &= (U+2E_c) \niu \nid \\
&+ \left[\epsilon - 2E_c (\hat{n}-n_0)  + E_c \right] \nimp \\
&+ E_c (\hat{n}-n_0)^2.
\end{split}
\end{equation}
In the canonical ensemble we may replace $\hat{n}$ by $n$ in each charge sector.
Thus the effective $U$ increases by $2E_c$, the level is shifted by $-2E_c (n-n_0)+E_c$, i.e.
$\delta=\epsilon+U/2$ is shifted by $-2E_c(n-n_0)+2E_c = -2E_c[n-(n_0+1)]$,
and the energy shift term becomes a constant, $E_c (n-n_0)^2$.
We may thus eliminate the quadratic charge terms in the SC, while the
impurity terms are renormalized. This is convenient for implementation and permits the
reduction of the MPO representation to $8 \times 8$ matrices, however this form
is less physically transparent.

\section{Benchmark calculations}

We verified the implementation at $\Gamma=0$, $E_c>0$ against the exact solution for the model
without the impurity \cite{Richardson1966}, finding full agreement within numerical roundoff
errors for energies.

\begin{figure}[htbp]
\centering
\includegraphics[width=0.8\columnwidth]{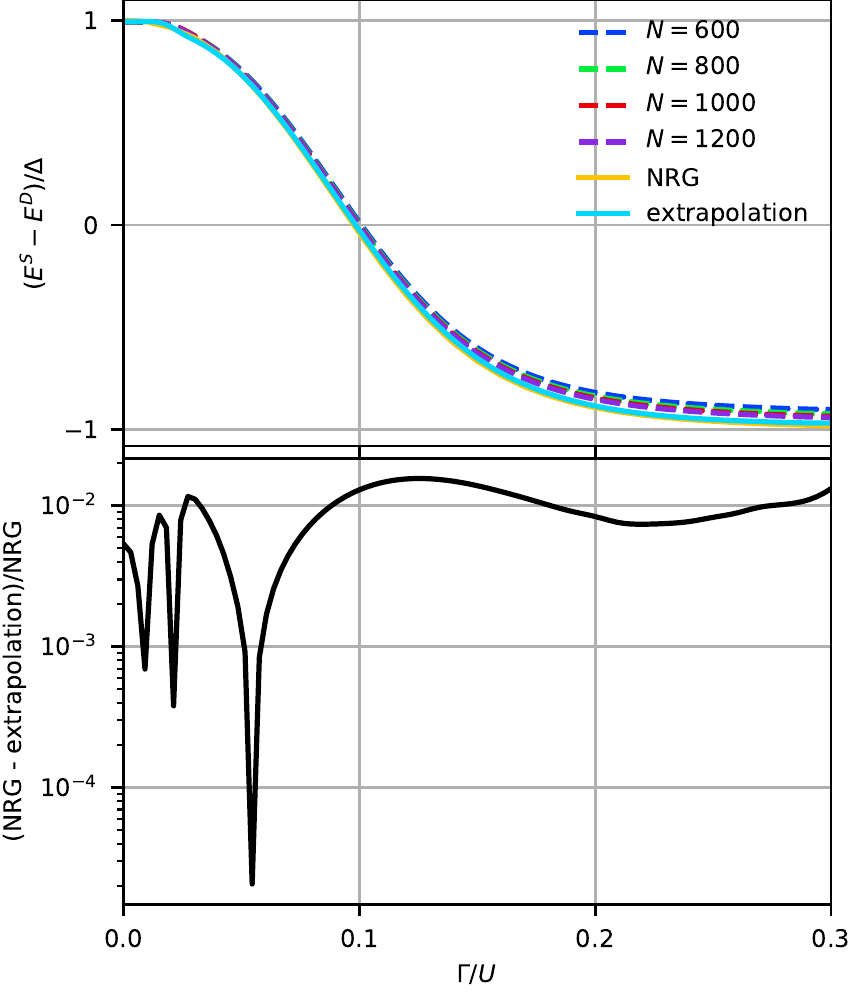}
\caption{Benchmark calculations for $E_c=0$.
(a) DMRG results for the YSR state energy $E_\mathrm{YSR}=E_S-E_D$ as a function of $\Gamma$ for a range of $N$ and the $N\to\infty$ extrapolation,
compared against the
results obtained using the numerical renormalization group (NRG) in the thermodynamic limit at the mean-field level.
(b) Difference between the extrapolated DMRG results and the NRG results.
}
\label{sup1}
\end{figure}

We verified the implementation at $\Gamma>0$, $E_c=0$ by comparing the results of numerical
renormalization group (NRG) calculations for mean-field BCS bath (parametrized by the BCS gap
value $\Delta$) and the DMRG calculations for interacting bath (parametrized by the pairing
coupling constant $\alpha$). The NRG calculations are performed in the thermodynamic limit but for
a logarithmically discretized bath, while the DMRG calculations are performed for large but finite
number of levels $N$. We remark that the NRG is not exact (due to truncation of states, which
leads to unavoidable systematic errors in addition to those due to logarithmic discretization),
while the DMRG has no systematic errors. Typical NRG errors for quantities such as excitation
energies are of the order of few percent \cite{resolution, errors}. We find that after the
$N\to\infty$ extrapolation of the DMRG results for the excitation energies of the subgap states,
we recover the NRG results within the expected error margin of a few percent, see Fig.~\ref{sup1}.
To obtain a mapping between $\Delta$ and $\alpha$, we performed the NRG calculations for a range
of $\Delta$ and selected the value where the agreement of the YSR excitation energies was optimal.
Since the value of the coupling constant $\alpha$ used in this work, $\alpha=0.23$, lies at the
boundary between weak and strong-coupling BCS regimes, this empirical approach is more reliable
than various analytical estimates for $\Delta$ as a function of $\alpha$.

We furthermore tested the implementation with all terms of the Hamiltonian, including the charging
terms with $E_c\neq 0$, against full diagonalisation on small clusters (up to $N=12$) using the
Lanczos method, finding full agreement within the numerical roundoff errors.

\end{document}